\documentclass[a4paper,11pt]{article}
\usepackage{graphicx} 

\pdfoutput=1 

\usepackage{jcappub} 

\usepackage[T1]{fontenc} 

\usepackage[utf8]{inputenc}
\usepackage{amsmath,mathtools}

\usepackage{tensor}
\usepackage{amssymb}
\usepackage{graphicx}
\usepackage{physics}
\usepackage{lipsum}  
\usepackage{bigints}
\usepackage{verbatim}

\usepackage{braket}
\usepackage{cancel}
\usepackage{booktabs}
\usepackage{color}
\usepackage{hyperref}
\usepackage{graphicx,subcaption}

\newcommand{\roughly}[1]{\mathrel{\raise.3ex\hbox{$#1$\kern-0.85em
\lower1ex\hbox{$\sim$}}}}
\newcommand{\lsim}{\roughly<}
\newcommand{\gsim}{\roughly>}

\newcommand{\mfa}{{\mathfrak a}}
\newcommand{\mfg}{{\mathfrak g}}
\newcommand{\bfg}{{\mathbf{g}}}
\newcommand{\bfx}{{\mathbf{x}}}

\newcommand{\cO}{{\cal O}}
\newcommand{\cU}{{\cal U}}
\newcommand{\MPL}{M_p}
\newcommand{\MEW}{M_{\scriptscriptstyle W}}
\newcommand{\ga}{{\mathbf{g_a}}}
\newcommand{\exd}{{\rm d}}
\newcommand{\pref}[1]{{(\ref{#1})}}
\newcommand{\CDM}{{\ssC}}
\newcommand{\B}{{\ssB}}

\newcommand{\nn}{\nonumber}

\newcommand{\ssB}{{\scriptscriptstyle B}}
\newcommand{\ssC}{{\scriptscriptstyle C}}
\newcommand{\ssT}{{\scriptscriptstyle T}}

\begin{document}

\title{CMB Implications of Multi-field Axio-dilaton Cosmology}

\author[a]{Adam Smith,}
\affiliation[a]{School of Mathematics and Statistics, University of Sheffield
}
\author[b,c,d]{Maria Mylova,}
\affiliation[b]{Kavli Institute for the Physics and Mathematics of the Universe (WPI)
}
\affiliation[c]{Science Education, Ewha Womans University
}
\affiliation[d]{Perimeter Institute for Theoretical Physics
}
\author[e]{Philippe Brax,}
\affiliation[e]{Institut de Physique Th\'eorique, Universit\'e Paris-Saclay
}

\author[a]{Carsten van de Bruck,}

\author[d,f,g]{C.P.~Burgess}

\affiliation[f]{Department of Physics \& Astronomy, McMaster University
}
\affiliation[g]{School of Theoretical Physics, Dublin Institute for Advanced Studies
}
\author[h,i]{and Anne-Christine Davis}
\affiliation[h]{DAMTP, University of Cambridge
}
\affiliation[i]{Kavli Institute of Cosmology (KICC), University of Cambridge
}

\emailAdd{asmith69@sheffield.ac.uk}
\emailAdd{mylova@g.ecc.u-tokyo.ac.jp}
\emailAdd{philippe.brax@ipht.fr}
\emailAdd{c.vandebruck@sheffield.ac.uk}
\emailAdd{cburgess@perimeterinstitute.ca}
\emailAdd{ad107@cam.ac.uk}

\abstract{

Axio-dilaton models are among the simplest scalar-tensor theories that contain the two-derivative interactions that naturally compete at low energies with the two-derivative interactions of General Relativity. Such models  are well-motivated as 
the low energy fields arising from string theory compactification. We summarize these motivations and compute their cosmological evolution, in which the dilaton acts as dark energy and its evolution provides a framework for dynamically evolving particle masses. The derivative axion-dilaton couplings play an important role in the success of these cosmologies. We derive the equations for fluctuations needed to study their 
implications for the CMB anisotropy, matter spectra and structure growth. We use a modified Boltzmann code to study in detail four benchmark parameter choices, including the vanilla Yoga model, and identify couplings that give viable cosmologies, including some with surprisingly large matter-scalar interactions. 
The axion has negligible potential for most of the cosmologies we consider but we also examine a simplified model for which the axion potential plays a role, using axion-matter couplings motivated by phenomenological screening considerations. We find such choices can also lead to viable cosmologies. 

}

\maketitle

\section{Introduction}

Cosmology is a Tale of Two Cities. It is the best of times (precision measurements spectacularly constrain our understanding of the recent universe); it is the worst of times (the concordance $\Lambda$CDM model reigns supreme). It is the epoch of belief (General Relativity -- GR -- cannot be improved upon); it is the epoch of incredulity (modifications to GR are everywhere). It is the season of light (cosmic microwave backround -- CMB -- observations get better and better); it is the season of darkness (95\% of the universal energy content consists of unknown Dark Matter -- DM -- and Dark Energy -- DE). It is the Spring of hope (small tensions in the data might undermine the cosmological consensus); it is the Winter of despair (systematic error is hard to quantify). We have everything before us (a cosmological constant describes well the evidence for Dark Energy); we have nothing before us (UV physics seems unable to give the observed cosmological constant in a technically natural way). Such times invite new ideas but ruthlessly cull those that do not measure up. 

In this paper we compute the cosmological implications of axio-dilaton scalar-tensor models, whose precise definition is given in \S\ref{MultipleScalars}. This section also argues them to define a broad class of minimal yet well-motivated two-field scalar-tensor theories whose broader implications for testing gravity are largely unexplored. For these models we here compute the evolution of both background fields and the linearized perturbations about these backgrounds required for computing CMB temperature-temperature correlations and the growth of large-scale structure. 

These models are well-motivated in two independent ways. First, two fields are the minimum number that allow the two-derivative sigma-model self-interactions that general power-counting arguments \cite{Burgess:2009ea, Adshead:2017srh} show compete most efficiently with the two-derivative interactions of GR at very low energies. Because both GR and the sigma-model interactions share the same number of derivatives they are able to compete at low energies without undermining the low-energy approximation on which the use of semiclassical methods in gravity ultimately relies \cite{Burgess:2003jk, Burgess:2020tbq}. It is because these interactions require at least two scalars that single-scalar models are usually driven to study higher-derivative interactions like those arising within the Horndeski program \cite{Horndeski:2024sjk} (at least once the dangerous zero-derivative interactions of the scalar potential are suppressed using shift symmetries).  

 Axio-dilaton models are also well-motivated because they encode features that are not unusual from the point of view of physics at much higher energies. Axio-dilatons are very commonly found amongst the low-energy fields in string vacua for robust symmetry reasons -- they capture the low-energy pseudo-Goldstone boson physics arising from the interplay of spontaneously broken internal symmetries (the axion) and approximate scaling symmetries (the dilaton) that are generic in string compactifications \cite{Burgess:2020qsc}. This makes their presence at low energies likely more robust than it would be if it only relied on conjectures about what is possible in principle at very high energies. Axio-dilatons would also be expected to be generic occupants of the Dark Sector in the (not unlikely  \cite{Burgess:2021juk}) event that the gravity sector should prove to be more supersymmetric than is the low-energy particle physics sector we explore in collider experiments. Finally, they also play a central role in attempts to understand Dark Energy in a technically natural way that exploit the nontrivial low-energy interplay between supersymmetry and scaling symmetries \cite{Burgess:2021obw}. 

\S\ref{MultipleScalars} fleshes out some of these motivations and defines more precisely the class of Lagrangians we use to define what we mean by axio-dilaton models, as well as the choices one is free to make within this class and some of the non-cosmological constraints they face. \S\ref{AxioDilatonCosmology} then specializes the field equations found in \S\ref{MultipleScalars} to cosmology, both for homogeneous background fields and for linearized fluctuations about these backgrounds. 

An important theme emerges from these equations: many of the implications for cosmology arise because particle masses are field-dependent (in Planck units) and so can vary in space and time as the fields themselves do. By doing so they provide a concrete dynamical framework within which to better test ideas (such as changing the value of the electron mass near recombination \cite{Sekiguchi:2020teg}) that have been proposed to help solve various cosmological tensions. Having a dynamical framework makes it possible to relate the values of masses at different epochs -- nucleosynthesis, recombination, during structure formation and at the current epoch -- and thereby find new ways to confront them with observations.

\S\ref{Section:Yoga Case} contains our main results, defining four benchmark choices of parameters whose properties we explore numerically using a modified version of the Cosmic Linear Anisotropy Solving System code (CLASS) \cite{Diego_Blas_2011}, which we employ to compute both the matter power spectrum and the CMB angular spectra. The coupling choices for these benchmarks are summarized for convenience in Table \ref{models}, together with a short summary of what we find for the viability of their cosmologies. As can be seen from the Table the most restrictive choice for couplings -- including in particular the cosmology described in detail in \cite{Burgess:2021obw} -- seems to be ruled out, but the other three variations on the theme are not (including examples with the large dilaton-matter couplings required in \cite{Burgess:2021obw} by the naturalness arguments for the Dark Energy). The difference between success and failure for the two models with strong dilaton-matter couplings hinges on the choices made for how the dilaton couples to Dark Matter.

None of the four of the benchmarks studied in \S\ref{Section:Yoga Case} include an axion potential and all assume the simplest possible linear coupling between axions and matter. These assumptions are relaxed in \S\ref{QuadraticAxionCouplingSection} which recomputes evolution for a simple choice for axion potential and for a quadratic matter-axion interaction that is motivated by phenomenological and screening considerations \cite{Brax:2023qyp}. We find that the inclusion of an axion potential makes the cosmologies easier to work and based on this we hope to explore these directions more systematically in future.

\begin{table}
\begin{center}
\noindent\makebox[\textwidth]{%
\begin{tabular}{ |l|l|p{6.5cm}|} 
\hline

Model & Features & Comment on background evolution \\

\hline

Universal Yoga Model & ${\bf g} = -\frac{\zeta}{2}$, $\zeta = \sqrt{\frac{2}{3}} $ & Likely not viable:  axion not sufficiently sourced to keep dilaton in local mini\-mum; larger axion couplings drain baryon density significantly.\\

\hline 

Yoga with opposite coupling & ${\bf g}_{\bf B} = -\frac{\zeta}{2}$, ${\bf g}_{\rm C} = -\frac{{\bf g}_B}{x}$ & Realistic background evolution for ~~~$1\lesssim x \lesssim 13$. \\ 

\hline

Reduced $\zeta$ & ${\bf g} = -\frac{\zeta}{2}$, $\zeta = x $ & Realistic background evolution for ~~~$x\lesssim 0.1 $. \\ 

\hline

General dilaton-matter coupling & ${\bf g} = -x$, $\zeta = \sqrt{\frac{2}{3}} $ & Realistic background evolution for ~~$x\lesssim 0.05 $. \\

\hline 
\end{tabular}
}
\end{center}
\caption{{\small Summary of the axiodilaton models considered in this paper. $\mathbf{g}_\ssB$ is the Brans-Dicke coupling strength between the dilaton and baryonic matter, $\mathbf{g}_\ssC$ is its analog for the dilaton-DM coupling and when these are equal they are both denoted simply $\mathbf{g}$ with no subscript. $\zeta$ is the strength of the kinetic axion-dilaton coupling $(\partial \chi)^2 + W^2 \, (\partial \mfa)^2$, where $W(\chi) = e^{-\zeta\chi}$, in (\ref{full action}).}}
\label{models}
\end{table}

Some conclusions are briefly summarized in \S\ref{Conclusions}.

\section{Multiple scalar motivations}
\label{MultipleScalars}

Our focus is on cosmological tests for multi-field scalar-tensor theories, so we first pause to define the models and motivate why they are interesting. Scalar fields are among the simplest possible deviations from GR with observable consequences, and as a result their implications for late-time cosmology have been widely studied. 

Most of such work -- with some exceptions, see {\it e.g.}~\cite{Amendola:1999dr, Hwang:2001fb, Malik:2002jb, Amendola:2014kwa, CarrilloGonzalez:2017cll, Johnson:2020gzn, Sa:2021eft, Eskilt:2022zky, vandeBruck:2022xbk, Poulot:2024sex} -- specialises when doing so to the case of a single scalar field (such as a Brans-Dicke field or an axion). Indeed, naturalness issues make this restriction to single fields seem reasonable: if light scalar fields are difficult to obtain from UV physics then we should be lucky to find even one of them relevant to late-time cosmology. This restriction to single scalar fields -- and the introduction of approximate shift symmetries to ensure they remain light -- in turn motivates the widespread study of higher derivative ({\it e.g.} `Horndeski') interactions: for single scalar fields these are the only scalar-tensor self-interactions consistent with shift symmetries.

\subsection{Multiple scalars}

But this focus on single-field models comes with its own problems. The first problem is with the higher derivative interactions to which they lead. These interactions are a problem whenever they compete with GR because it is ultimately the derivative expansion that justifies using classical reasoning in cosmology (for reviews see \cite{Burgess:2003jk, Burgess:2020tbq}); so higher-derivative interactions can only compete with the two-derivative interactions of GR when the entire semi-classical approximation is starting to fail \cite{Burgess:2009ea, Adshead:2017srh} (DBI models provide one of the few known exceptions to this general statement\footnote{Notice that the dangerous wavelengths from this point of view are those much shorter than the curvature radius and so these conclusions are more restrictive than the ones often quoted in cosmology, which study scales for which the background curvatures can be competitive (see {\it e.g.}~\cite{deRham:2014wfa}).} \cite{Babic:2019ify}).

The general power-counting arguments of \cite{Burgess:2009ea, Adshead:2017srh} show (unsurprisingly) that it is two-derivative `$\sigma$-model' interactions\footnote{We use a positive signature metric and MTW curvature conventions and relate the Planck mass to Newton's constant by $\MPL^{-2} = 8\pi G_{\rm N}$.}
\begin{equation} \label{SigmaL2}
   {\cal L}_{2-{\rm deriv}} = \tfrac12 \, \MPL^2 \sqrt{-g} \Bigl[ R - {\cal G}_{ab}(\phi) \, \partial_\mu \phi^a \, \partial^\mu \phi^b \Bigr] \,,
\end{equation}
amongst $N$ scalars and the metric that dominate at low energies once shift symmetries are used to suppress the dangerous zero-derivative interactions of the scalar potential. (We note in passing that the shift symmetries $\delta \phi^a = \xi^a(\phi)$ used to suppress $V(\phi)$ can be completely consistent with ${\cal G}_{ab}$ being $\phi$-dependent.) It is the $\sigma$-model interactions contained within the $\phi$-dependence of ${\cal G}_{ab}$ that want to compete at low energies with the two-derivative interactions of GR. The positive definite matrix of functions ${\cal G}_{ab}(\phi)$ transform under field redefinitions as would a symmetric tensor on the $N$-dimensional target space spanned by the dimensionless fields $\phi^a$ and so can be viewed as a metric on the target space. 

It turns out that interactions like (\ref{SigmaL2}) have no physical effect (beyond the minimal coupling to gravity) when the target-space geometry associated with ${\cal G}_{ab}$ is flat because it is then always possible to arrange ${\cal G}_{ab} = \delta_{ab}$ by performing an appropriate field redefinition. That is why the ${\cal G}_{ab}$ interactions never play a role for single-scalar models: all one-dimensional target space manifolds are flat. A focus on single-field models accidentally removes the possibility of two-derivative $\sigma$-model interactions, potentially missing the important class of couplings that is expected naturally to compete with GR at the low energies accessible to cosmology.

Having multiple light scalars can also be just as natural as having just one, at least in a world with a small vacuum energy \cite{Albrecht_2002}. This is because the natural extension of (\ref{SigmaL2}) to include a scalar mass adds the small shift-symmetry breaking term
\begin{equation} \label{SigmaL0}
    {\cal L}_{\rm pot} = - \sqrt{-g} \; v^4 {\cal S}(\phi) \,,
\end{equation}
where $v$ is a mass scale and ${\cal S}(\phi)$ is an order-unity dimensionless function of the dimensionless $\phi^a$. The cosmological constant problem asks why $v$ is as small as is currently observed, but once this has been answered the mass predicted for $\phi^a$ by combining (\ref{SigmaL2}) and (\ref{SigmaL0}) is generically $v^2/\MPL$ provided only that ${\cal S}(\phi)$ and its derivatives are all order unity. {\it Any} gravitationally coupled scalars would therefore be expected to have a Hubble-sized mass in a world where the potential is small. The cosmological constant problem itself (why is $v$ small) is the only naturalness problem in such a world and the existence of multiple light gravitationally coupled scalars need not be {\it additionally} unnatural.

\subsection{Well-motivated two-field examples}

Given the motivation to explore 2-derivative $\sigma$-model interactions amongst multiple light scalars, two-scalar models are natural places to start because they have the minimal number needed to allow their existence. The freedom to perform field redefinitions ensures that the most general two-dimensional target space metric can always be locally written in terms of a single nonvanishing function $Z(\phi,\psi)$, with
\begin{equation} \label{2FieldL}
  {\cal L}_{2-deriv} = - \tfrac12  \MPL^2 \, \sqrt{-g} \; Z^2(\phi,\psi) \, \Bigl[ (\partial \phi)^2 + (\partial \psi)^2 \Bigr] \,.
\end{equation}
This, together with the choices for scalar potential and for how the scalars couple to matter, still leaves a large class of models to be explored. To make further progress it is helpful to narrow the class of models under consideration still further. The axio-dilaton class of models provides a well-motivated subclass worthy of more detailed study. 

\subsubsection{Axio-dilatons}
Axio-dilaton models are defined by two assumptions: target-space axion shift symmetry and dilaton-like matter couplings, each of which is discussed in more detail below. 

The axionic shift symmetry assumes ${\cal G}_{ab}$ is invariant under constant shifts of one of the two fields (which we call\footnote{We here follow that part of the literature that calls any axion-like Particle -- or ALP -- an axion and thereby alienate that other part of the literature that reserves the word axion for the specific QCD axion whose potential existence was recognized \cite{Weinberg:1977ma, Wilczek:1979hc} some time ago to arise in solutions to the strong-CP problem \cite{Peccei:1977hh, Peccei:1977ur}.} an `axion' and so denote by $\mathfrak{a}$): $\mfa \to \mfa + c$ for constant $c$. This makes the target-space metric `axisymmetric' inasmuch as it is independent of $\mfa$. In this case the function $Z$ is independent of $\mfa$ and (\ref{2FieldL}) can be written in either of the following two equivalent ways:
\begin{equation}
  {\cal L}_{2-deriv} = - \tfrac12  \MPL^2 \, \sqrt{-g} \; Z^2(\phi) \, \Bigl[ (\partial \phi)^2 + (\partial \mfa)^2 \Bigr] = - \tfrac12  \MPL^2 \, \sqrt{-g} \; \Bigl[ (\partial \chi)^2 + W^2(\chi) (\partial \mfa)^2 \Bigr] \,,
\end{equation}
where $Z(\phi) \exd \phi = \exd \chi$ and $W(\chi) = Z[\phi(\chi)]$. 

The second assumption is that the field $\phi$ couples to matter like a dilaton; {\it i.e.}~like a pseudo-Goldstone boson for an approximate scaling symmetry. In practice, we take this to mean that $\phi$ couples to matter as does a Brans-Dicke scalar: with matter Lagrangian density 
\begin{equation} \label{MatterLAnsatz}
    {\cal L}_m = {\cal L}_m[\tilde g_{\mu\nu},\mfa,\psi] \,,
\end{equation}
where $\psi$ collectively denotes any matter fields and the `Jordan frame' metric is defined by\footnote{The choice \pref{JFvsEF} with exponential function $C(\chi)$ in Einstein frame is equivalent to -- and much more useful than -- the more traditional Brans-Dicke definition \cite{Brans:1961sx}. The coupling $\bfg$ defined in \pref{JFvsEF} is related to the traditional Brans-Dicke parameter $\omega$ by $2\bfg^2 = (3+2\omega)^{-1}$.} 
\begin{equation} \label{JFvsEF}
  \tilde g_{\mu\nu} = C^2(\chi) \, g_{\mu\nu} \quad \hbox{where} \quad C(\chi) = e^{\mathbf{g} \chi} \,,
\end{equation}
for some coupling constant $\mathbf{g}$. Later we consider variations on this theme that relax the assumption that the dilaton couples universally to all matter fields, allowing baryons and dark matter to couple to different Jordan-frame metrics of the form given in \pref{JFvsEF} but with different couplings $\mathbf{g} _\ssB \neq \mathbf{g}_\CDM$.

Brans-Dicke couplings arise very commonly in UV completions of gravity, with $\phi$ and $\mfa$ often appearing as the real and imaginary parts of a complex field that is part of a supersymmetric multiplet (making the dilaton a `saxion': the scalar super-partner of an axion). A commonly arising example within such constructions has target space
\begin{equation} \label{SL2RTarget}
    {\cal G}_{ab} \partial_\mu \phi^a \, \partial^\mu \phi^b = \frac{(\partial \phi)^2 + (\partial \mfa)^2}{\zeta^2\phi^2 } \, ,
\end{equation}
for a constant $\zeta$, and so 
\begin{equation}
    Z(\phi) = \frac{1}{\zeta  \phi} \qq{which implies}
    \phi = e^{\zeta \chi} \qq{and}
    W(\chi) = \frac{1}{\zeta} \, e^{-\zeta \chi} \,.
\end{equation}
In this case the metric ${\cal G}_{ab}$ is the $SL(2,R)$-invariant metric on the hyperbolic upper-half plane so we call this the $SL(2,R)$-invariant axio-dilaton (or the Angle-Saxion model).

\subsubsection{Yoga models}
\label{yoga models}

Yoga models \cite{Burgess:2021obw} are a specific subset of the axio-dilaton class of two-scalar theories that is motivated by an approach to the cosmological constant problem based on a `natural relaxation' mechanism (hence the name). The relaxation mechanism assumes that a scaling symmetry and supersymmetry survive to low energies {\it in the dark sector}, such as can arise as accidental approximate symmetries within the gravitationally coupled sector within string compactifications \cite{Burgess:2020qsc}. 

In these models the fields $\phi$ and $\mfa$ are the real and imaginary parts of a complex scalar and for the present purposes what is important is that the underlying symmetries predict $Z^2 = \frac32/\phi^2$ and so $\phi = e^{\zeta \chi}$ with $\zeta = \sqrt{\frac23}$. The scaling symmetry requires ordinary baryons to couple to the dilaton with a Jordan-frame metric $\tilde g_{\mu\nu} = g_{\mu\nu}/\phi$ and so predicts the Higgs expectation value (and so also non-neutrino particle masses) to be proportional to $\phi^{-1/2}$ and implies a Brans-Dicke coupling to matter of size $\mathbf{g} _\ssB  = - \frac12 \zeta = -\frac{1}{\sqrt 6}$. 

These models by design predict a scalar potential for the dilaton of the form
\begin{equation} \label{YogaPotential}
   V_{\rm Yoga} = \frac{U \MPL^4}{\phi^4} = \MPL^4 \, U \, e^{-4\zeta \chi}\,,
\end{equation}
where $U$ is a function of $\ln \phi$ (or, equivalently, of $\chi$). The point of the models is that these structures can be technically natural. Because $U$ depends only logarithmically on $\phi$ it can easily allow minima for which $\phi$ is exponentially large. (The simplest examples choose $U$ to be quadratic in $\ln \phi$ with coefficients of order 50 in size.) With these choices $\chi$ can easily be minimized at a value $\chi_{\rm min} \sim 60$, in which case $\phi_{\rm min}$ is order $(\MPL/\MEW)^2 \sim 10^{28}$. For such values particle masses are predicted to be of order $m \sim \MPL/\sqrt{\phi_{\rm min}} \sim \MEW$ while $V_{\rm Yoga}(\phi_{\rm min})  \sim \MPL^4/\phi_{\rm min}^4 \sim (\MEW^2/\MPL)^4$, both of which have the right order of magnitude.\footnote{As ever, the devil is in the details and \cite{Burgess:2021obw} explores more precisely how small $V_{\rm Yoga}(\phi_{\rm min})$ can be.} As a bonus, if neutrino masses are generated by the dimension-5 Weinberg operator \cite{Weinberg:1979sa, Weinberg:1980bf} then they would have size $m_\nu \sim M_p/\phi_{\rm min} \sim \MEW^2/\MPL$ which is also successful in order of magnitude.

No prediction is made in these models for what the couplings of Dark Matter should be to $\phi$, since the model does not specify what the Dark Matter is. (For instance, if DM were primordial black holes - pBHs - then the no-hair theorem would predict the DM-$\chi$ coupling to be zero.) However it was also observed in \cite{Burgess:2021obw} that if Dark Matter were also to have masses proportional to $\phi^{-1/2}$ then the model might be able to realise the mechanism proposed in \cite{Sekiguchi:2020teg} to resolve the Hubble tension by changing the electron mass at recombination. (Indeed part of the motivation for the paper you are now reading is to explore more systematically how these models fare when compared with the CMB.) If this were the case then the Brans-Dicke coupling of $\chi$ to Dark Matter would also be $\mathbf{g}_\CDM = - \frac{1}{\sqrt6}$. For later purposes we call this specific form of the Yoga proposal the `Universal' Yoga model.  

\subsection{Potentials and non-cosmological constraints}
\label{NonCosmoConstraints}

Although the focus of the rest of this paper is cosmology it is worth remarking here on the various non-cosmological challenges that these models face. These involve constraints on the matter couplings of both the dilaton and the axion, which we briefly summarize here. These constraints also motivate some of the choices we make for the model's axion potential and matter-axion couplings.

\subsubsection{Dilaton constraints}

The couplings of a Brans-Dicke scalar are strongly constrained by tests of gravity within the solar system, such as by tests of Shapiro time delay by the Cassini probe \cite{Bertotti:2003rm}. For scalars with Compton wavelengths much larger than typical solar system scales (such as are considered here) these require the scalar-matter coupling to satisfy $\bfg_\ssB \lesssim 10^{-3}$ in order to have been too small to have been detected. The simplest option when considering cosmology is simply to restrict $\bfg_\ssB$ to satisfy this bound. This in particular would exclude the Yoga models, which predict $\mfg_\ssB = - 1/\sqrt6$. 

We do not impose this condition here and instead ask whether viable cosmologies can exist even with Yoga-sized $\cO(1)$ dilaton couplings. We do so because it need not be true that the coupling $\bfg_\ssB$ appearing in cosmology is the same coupling that is constrained in non-cosmological tests. In theories where fields couple nonlinearly to one another and to matter it can happen that the dilaton coupling relevant to a macroscopic object (like the Sun, in solar system constraints) is not simply the sum of the dilaton coupling to each of its constituent particles. When macroscopic couplings are much smaller than the sum of microscopic couplings the force is said to be `screened' and a variety of mechanisms have been found \cite{Hinterbichler:2010es, Khoury:2003aq, Brax:2010gi} to accomplish screening in scalar-tensor theories (including explorations within the axio-dilaton class of models considered here \cite{Burgess:2021qti, Brax:2022vlf, Brax:2023qyp}). 

The search for screening mechanisms in axiodilaton models remains young and so we do not restrict to specific models for which a detailed screening mechanism is known, instead exploring parameter choices that seem broadly promising for screening.

\subsubsection{Axion constraints}

The axion's couplings to matter are also a potential worry since these are also subject to a wealth of constraints \cite{ParticleDataGroup:2022pth}. In principle these constraints are more model-dependent than are those for the dilaton because there there is considerable latitude in choosing the axion potential and its couplings to matter.

A generic worry when $\phi$ is as large as chosen for the Yoga models is the size of the axion kinetic term \pref{SL2RTarget}. For any specific value $\chi = \bar \chi$ the canonically normalized axion field is $a_c = f \, \mfa$ where $f(\bar\chi) = M_p W(\bar\chi) = M_p/\bar\phi$ provides one definition of the axion decay constant. When $\bar\phi = \phi_{\rm min} \sim (\MPL/\MEW)^2$ the value predicted for $f$ then is of order $\MEW^2/\MPL$ (and so is of order the eV scale). This seems in immediate conflict with some axion constraints (such as cooling rates from red giant stars) that generally require the decay constant $F$ appearing in matter couplings to satisfy $F \gsim 10^9$ GeV. 

Whether this is really a problem proves to be a model-dependent issue because it is not always appropriate to identify the physical decay constant with the coefficient of the axion kinetic term (some explicit UV completions for axions for which this is true are given in \cite{Brax:2022vlf}). We therefore treat the quantity $f$ defined by the kinetic term and $F$ defined by the matter couplings as independent. In what follows we typically find $F = f/\gamma$ for some parameter $\gamma$, whose value is then checked to be sufficiently small. (In the examples we explore $\gamma \sim (\MEW/\MPL)^2$ and so $F \sim M_p$.)

\subsubsection{Scalar potentials}

Up to this point little has been said -- apart from \pref{YogaPotential} -- about the nonderivative couplings of $\phi$ and $\mfa$. In the spirit of \cite{Burgess:2021obw} for most of our analysis we neglect any axion-dependence of the scalar potential, assuming \pref{YogaPotential} provides the dominant contribution. Part of the reason for this is to see whether interesting cosmology can arise when both $\phi$ and $\mfa$ are free to evolve even relatively recently.

But again following \cite{Burgess:2021obw}, and motivated by the screening discussion above, we do allow non-derivative couplings between $\mfa$ and matter. The idea behind the screening constructions is: if matter can couple to both the axion and the dilaton then the nonlinear interactions amongst these two fields can reduce the dilaton field outside a macroscopic source like the Sun. (See \cite{Brax:2023qyp} for the most promising proposal along these lines so far.) For concreteness' sake we here regard this to arise microscopically as an axion-dependence to particle masses
\begin{equation} \label{AxionDependentMass}
    m_f = m_{f0}(\chi) \,\cU_f(\mfa) ,
\end{equation}
for some choice of functions $\cU_f(\mfa)$. 

Motivated by the QCD axion one might also expect symmetry-breaking matter-axion interactions with matter to come hand in hand with an axion dependence of the scalar potential in vacuum. This indeed can -- but need not -- arise depending on the strength of the couplings that lead to nontrivial $\cU_f(\mfa)$. In the specific case of Yoga models this would be expected to imply that the function $U$ appearing in \pref{YogaPotential} acquires a dependence on $\mfa$. 

The screening mechanism proposed in \cite{Brax:2023qyp} relies on $\mfa$ taking values near the minimum of $\cU(\mfa)$ within ordinary matter and on this minimum differing from the minimum of the axion's vacuum potential. Axion constraints, like those coming from energy loss from hot stars, also are less stringent if $\mfa$ sits near the minimum of $\cU(\mfa)$ within the radiating objects. We therefore in \S\ref{QuadraticAxionCouplingSection} make a preliminary attempt to assess the viability of cosmology consistent with these types of interactions, assuming a simplified scalar potential $V = V(\mfa) + V_{\rm Yoga}(\phi)$ when doing so (rather than having $U$ appearing within $V_{\rm Yoga}$ also be $\mfa$-dependent). 

We find sensible cosmologies can exist with axion potentials of this type, motivating a more detailed study of more realistic potentials. We defer this more systematic exploration of axion-dependent potentials to future work, but with an eye to this work we derive our analytic expressions for the axio-dilaton cosmology in the next section assuming a completely general dependence of the potential on the axion and dilaton.

\section{Axio-dilaton cosmology}
\label{AxioDilatonCosmology}

This section derives the field equations needed to describe both background cosmologies and fluctuations around these for two-scalar models of the axio-dilaton class.

\subsection{Field equations}

The effective field theory (EFT) whose equations we wish to solve has the action 
\begin{equation}\label{full action}
    S=\frac{1}{2}\int \dd^4x\sqrt{-g}\Bigl\{ \MPL ^2 g^{\mu\nu}\Bigl[ R_{\mu\nu} -\partial_\mu \chi \, \partial_\nu\chi-W^2(\chi) \, \partial_\mu \mfa \, \partial_\nu \mfa\Bigr] -2 \, V\left(\chi,\mfa \right)\Bigr\} + \mathcal{L}_m \, ,
\end{equation}
where $\mathcal{L}_{m}$ parameterizes the contribution from the standard model and dark matter. The dilaton $\chi$ and the axion $\mfa$ are kinetically coupled to each other via the function $W(\chi)$ and in this section we allow the potential $V$ to depend quite generally on both scalar fields. The matter action is assumed to have the form given in \pref{MatterLAnsatz}, with matter coupling to $\chi$ only through a minimal coupling to a Jordan-frame metric 
\begin{equation}
  \tilde g_{\mu\nu}=C^{2}(\chi) \, g_{\mu\nu} \,,
\end{equation}
though its couplings to ${\mfa}$ can be more general (and are specified in more detail below).

Varying the action with respect to the metric gives us the modified Einstein equations
\begin{align}
\label{einstein eq}
 & G_{\mu\nu}-\left(\partial_\mu \chi \partial_\nu \chi - \frac{1}{2} g_{\mu\nu} g^{\rho\sigma} \partial_\rho \chi \partial_\sigma \chi \right) \nonumber
\\& \qquad\qquad\qquad -
W^2 \left( \partial_\mu \mfa \partial_\nu \mfa - \frac{1}{2} g_{\mu\nu} g^{\rho\sigma} \partial_\rho \mfa \partial_\sigma \mfa \right) 
+\frac{1}{\MPL ^2} \Bigl( g_{\mu\nu} V - T_{\mu\nu} \Bigr) = 0 \, ,  
\end{align}
where we define the (Einstein-frame) energy-momentum tensor as 
%
\begin{equation}
    T^{\mu\nu}=\frac{2}{\sqrt{-g}}\frac{\partial\mathcal{L}_m}{\partial g_{\mu\nu}} \,.
\end{equation}
The dilaton field equation similarly is
\begin{equation}
\label{dilaton eom}
    \Box\chi -W \, W,_\chi\partial_\mu\mfa\partial^\mu\mfa-\frac{V,_\chi}{\MPL ^2}=-\frac{\mathbf{g}T}{\MPL ^2}  \, ,
\end{equation}
where $T := g_{\mu\nu} T^{\mu\nu}$ and the dilaton-matter coupling is $\mathbf{g} := C_{,\chi}/C$. In practice we choose $C = \exp({\mathbf{g} \chi})$ with $\mathbf{g}$ a constant, in which case $\chi$ couples to matter in the same way as would a Brans-Dicke scalar. Later sections entertain the option of having the value for the dilaton coupling $\mathbf{g}$ differ for ordinary matter and for dark matter. 

Finally, the axion field equation is
\begin{equation}
\label{axion eom}
 \Box\mfa + \frac{2W,_\chi}{W} \, \partial_\mu\chi\partial^\mu\mfa-\frac{V,_\mfa}{W^2\MPL ^2}=-\frac{\mathbf{{\cal J}}}{W^2\MPL ^2}  \,,
\end{equation}
which denotes $\partial {\cal L}_m/\partial \mfa = \sqrt{-g} \; {\cal J}$. 

In these expressions ordinary (and dark) matter contribute through the stress-energy tensor $T_{\mu\nu}$ and the axion source function ${\cal J}$. Their contribution to the stress energy is (as usual for cosmological applications) obtained by writing $T_{\mu\nu}$ as the sum of perfect-fluid contributions 
\begin{equation}
     T_{(f)}^{\mu\nu}  = (\rho_f + p_f) u^\mu_f u^\nu_f + p_f g^{\mu\nu} \,,
\end{equation}
with $f = B,C$ or $R$ (for baryons, cold dark matter and radiation, respectively). Each fluid has a local 4-velocity, $u^\mu_f$, energy density, $\rho_f$, and pressure, $p_f$, related by an appropriate equation of state. After recombination stress energy is separately conserved for each of these fluids but because electromagnetic interactions allow baryons and photons to equilibrate they can exchange energy before recombination and so are not then separately conserved.  

In what follows we assume the microscopic axion-matter couplings arise as an axion-dependence to particle masses, as in \pref{AxionDependentMass}. If so the axion source evaluated in a nonrelativistic fluid can be written 
\begin{equation} \label{cJvsRhoB0}
     {\cal J} = -\sum_f \left( \frac{\partial m_f}{\partial \mfa} \right) n_f  \, ,
\end{equation} 
where $n_f$ is the particle's local number density and the sum runs over all species of particles that share the axion-dependent contribution $\cU(\mfa)$. In the applications to follow we imagine the axions to couple only to nonrelativistic species, in which case
\begin{equation} \label{cJvsRhoB}
     {\cal J} = -\sum_f \ga_f \rho_f \simeq -\ga \rho_\ssB\, ,
\end{equation} 
where $\rho_f \simeq m_f n_f$ is the particle energy density and $\ga_f := \cU_f'/\cU_f$. The final approximate equality assumes the summed energy density is dominated by the energy density of baryons.

Using this in \pref{axion eom} shows that the axion moves as if governed by an effective potential 
\begin{equation} \label{VeffMat}
V_{\rm eff}(\mfa, \rho_\ssB)\equiv V(\mfa) + \sum_f m_f(\mfa) \, n_f \simeq V(\mfa) + \rho_\ssB (\mfa) \,, 
\end{equation}
and so in particular can be interpreted to have a density-dependent effective mass 
\begin{equation} \label{meffdef}
    W^2\MPL^2 m_{(\rm eff)}^2 (\mfa, \rho_\ssB) = V_{\mfa\mfa}(\mfa) + \sum_f m_{f0} \, \cU_{\mfa\mfa}(\mfa) \, n_f 
    \simeq  V_{\mfa\mfa}(\mfa) + \left( \frac{\cU_{\mfa\mfa}}{\cU} \right)\, \rho_\ssB  .
\end{equation} 
For axions both $V$ and $\cU$ are usually periodic functions of $\mfa$, but in what follows we (like much of the literature) evolve using simpler approximate forms for these functions (that are specified in more detail in \S\ref{Section:Yoga Case}) and \S\ref{QuadraticAxionCouplingSection}. 

Notice that it is the decay constant $f = M_p W$ that appears on the left-hand side of \pref{meffdef}, whereas eq.~\pref{cJvsRhoB} shows that the decay constant $F$ setting the physical size of matter interactions is given by $F^{-1} = \ga/f$, which can be very different. It is the quantity $F$ that is subject to non-cosmological constraints like those coming from cooling rates from red giants and supernovae.

The equations governing the evolution of the components of the cosmic fluid are obtained from the flow of conserved quantities ({\it e.g.}~stress-energy and baryon number). The ability of the cosmic fluids to exchange energy with the scalar fields implies that the stress energy for each fluid component is not separately conserved (even in the absence of the equilibrating baryon-photon interactions mentioned above). That is, although the total energy-momentum tensor is covariantly conserved (i.e. $ \nabla_\mu T^\mu_{ \ \nu} =0$), at lowest order in a derivative expansion the exchange of energy amongst the fluid components and between the fluids and the scalars can be expressed as
\begin{equation}\label{T covar deriv}
    \nabla_\mu T_{(f)}^{\mu\nu} = J^\nu_{(f){\rm eq}} + Q^\chi_{(f)} \, g^{\nu\lambda} \partial_\lambda \chi + Q^\mfa_{(f)} \, g^{\nu\lambda} \partial_\lambda \mfa \, ,
\end{equation}
where the coefficients $J^\nu_{(f){\rm eq}}$ and $Q^i_{(f)}$ are determined as follows. 

First, the coefficients $Q^i_{(f)}$ can be read off from the coupling of the fluid component $f$ to each scalar field by demanding that the energy flow out of the fluid exactly cancel the energy flow into the scalar field, as dictated by the field equations \pref{dilaton eom} and \pref{axion eom}. For instance the dilaton field equation \pref{dilaton eom} suggests $Q_{(B)}^\chi = -\textbf{g}_\ssB \, \rho_\ssB$ for its coupling to baryons. Assuming a similar Brans-Dicke style coupling to cold dark matter similarly implies $Q_{(\CDM)}^\chi = -\mathbf{g}_\CDM \, \rho_\CDM$. The Universal Yoga model of \S\ref{yoga models} assumes $\mathbf{g}_\CDM = \textbf{g} _\ssB $, but in what follows we also explore other options because the Yoga mechanism is mute on the nature of dark matter and so does not require a particular choice for $\mathbf{g}_\CDM$. The axion field equation \pref{axion eom} and \pref{cJvsRhoB} similarly suggests $Q_{(\ssB)}^\mfa = -\mathbf{g_a} \, \rho_{\ssB}$. Because we assume no axion coupling to dark matter we choose $Q_{(\CDM)}^\mfa = 0$. 

Next, $J^\nu_{(f){\rm eq}}$ describing photon-baryon exchange is given by the usual expression involving the Thomson cross section (see for instance \cite{Ma:1995ey}) in which we use the dilaton-dependent electron mass $m_e(\chi) = m_{e0}/\phi^{1/2}$ predicted by the axio-dilaton framework. In practice we implement this by evolving the linearized photon and baryon fluids numerically using CLASS \cite{Diego_Blas_2011}, adjusted to include both a field-dependent Thomson scattering cross section and to include the fluid-scalar energy exchange terms $Q^i_{(f)}$ mentioned above. We drop the contributions from $J^\nu_{(f){\rm eq}}$ when analytically exploring post-recombination structure formation in later sections.

We next use these equations to describe linearized fluctuations about homogeneous and isotropic cosmological background solutions for this system.

\subsection{Background dynamics}

We assume a background describing a homogeneous and isotropic universe, and so choose a Friedmann–Lemaître–Robertson–Walker (FLRW) metric,
\begin{equation}
    {\rm d}s^2 = a^2\left[-{\rm d}\eta^2 + \delta_{ij} \, {\rm d}x^i {\rm d}x^j\right] \, ,
\end{equation}
where $a=a(\eta)$ and we assume flat spatial slices. Homogeneity and isotropy also imply space-independent background scalar field configurations, $\bar\chi = \bar \chi(\eta)$ and $\bar\mfa = \bar\mfa(\eta)$, so the equations describing the evolution of the background simplify to
\begin{align}
   & \mathcal{H}^2 = \frac{1}{3\MPL ^2}\left[\left(\frac{\Bar{\chi}'^2}{2}+\frac{W^2\Bar{\mfa}'^2}{2}\right)\MPL ^2+a^2V+a^2\Bar{\rho}\right] \, ,\nn
    \\
    \label{friedmann dilaon}
    & \Bar{\chi}''+2\mathcal{H}\Bar{\chi}'-WW,_\chi\Bar{\mfa}'^2+\frac{a^2}{\MPL ^2}\Bigl( V,_{\chi} +\mathbf{g} _\ssB  \, \Bar{\rho}_\ssB +\mathbf{g}_\CDM \, \Bar{\rho}_\CDM \Bigr)=0 \, ,
    \\
    \label{friedmann axion}
&\Bar{\mfa}'' +2 \mathcal{H}\Bar{\mfa}'+\frac{2W,_\chi}{W} \, \Bar{\mfa}'\Bar{\chi}'+\frac{a^2}{\MPL ^2 W^2}\Bigl( V,_\mfa +\mathbf{g_a} \,\rho_\ssB \Bigr) =0\, .\nn
\end{align}
where $\Bar{\rho} = \Bar{\rho}_\ssB + \Bar{\rho}_\CDM + \Bar{\rho}_{\rm rad}$, the Hubble expansion rate is $\mathcal{H} = a'/a$ and primes denote differentiation, $\dd /\dd \eta$, with respect to conformal time. Subscripts $_,\chi$ and ${_,\mfa}$ respectively denote differentiation with respect to the corresponding fields. 

Assuming no CDM-axion coupling the continuity equation \pref{T covar deriv} for the energy density of dark matter becomes
\begin{equation} \label{contequ CDM}
  \bar{\rho}'_\CDM+3\mathcal{H}\Bar{\rho}_\CDM =\mathbf{g}_\CDM\, \Bar{\rho}_\CDM \,\Bar{\chi}' \,.
\end{equation}
This equation has a simple physical interpretation: it describes an Einstein-frame energy density $\Bar{\rho}_\CDM = n_\CDM m_\CDM(\chi)$ with a number density $n_\CDM \propto a^{-3}$ (as appropriate for a fixed number of particles) and a dilaton-dependent mass $m_\CDM = \mathbf{m} \, \exp( \mathbf{g}_\CDM \, \chi)$. 

The conservation equation for the background energy in radiation is the standard one that implies $\bar\rho_{\rm rad} \propto a^{-4}$. Conservation of baryon number similarly ensures $n_\ssB \propto a^{-3}$. Conservation of energy for the baryon fluid is modified to\footnote{Notice that the equilibrating baryon-photon energy flow $J^\nu_{(\ssB){\rm eq}}$ is proportional to the difference between the baryon and photon expansion and so vanishes for the background evolution.}
\begin{equation}\label{contequ B}
 \bar{\rho}'_{\rm B}+3\mathcal{H}\Bar{\rho}_{\rm B}=\Bar{\rho}_{\rm B}\Bigl( \mathbf{g} _\ssB \, \Bar{\chi}' + \mathbf{g_a} \, \Bar{\mfa}'\Bigr) \,,
\end{equation}
which has an interpretation similar to the dark matter evolution equation, but with a particle mass that depends on both $\chi$ and $\mfa$.  

In summary, equations (\ref{friedmann dilaon}) through (\ref{contequ B}) describe the background evolution of the fields $\bar\chi$ and $\bar \mfa$ and how these exchange energy with the components of the cosmic fluid. We seek to explore the significance of this energy transfer for late-time cosmology and how this depends on the choices made for the parameters $\mathbf{g} _\ssB $, $\mathbf{g}_\CDM$, $\mathbf{g_a}$ and the kinetic coupling function $W$. 

\subsection{Perturbations}

In order to investigate the effects of axio-dilaton cosmologies on structure formation and the CMB, we now perturb the field equations to linear order. We work in the conformal-Newtonian gauge and assume only scalar perturbations, with metric
\begin{equation}
    {\rm d}s^2 = a^2(\eta)\Bigl[-(1+2\Phi){\rm d}\eta^2 + (1-2\Psi){\rm d}x^2 \Bigr] \, , 
\end{equation}
where $\eta$ again denotes conformal time. The components of the perturbed 4-velocity are denoted 
\begin{equation}\label{4VelocityExpansion}
u^0_f = \frac{1}{a} (1- \Phi)  \qq{and} u^i_f = \frac{ \partial^i v_f}{a} \, ,
\end{equation}
for each fluid, while the perturbed energy densities and scalar fields are
\begin{equation}
\rho_\ssB \equiv \bar\rho_\ssB (1+\delta_B) \, , \quad \rho_\CDM \equiv \bar\rho_\CDM (1+\delta_\CDM) \, , \quad \chi \equiv  \bar\chi + \delta\chi \qq{and}  \mfa \equiv\bar{\mfa}+\delta\mfa \, ,
\end{equation}
and we assume in the following that the anisotropic stress vanishes ($\Sigma_{ij} =0$). 

Going to Fourier space, the perturbed $(0,0)$, $(0,i)$  and $(i,j)$ components of the Einstein field equations (\ref{einstein eq}) respectively read
\begin{align}
\label{Perturbed Friedmann}
& k^2\Psi+3\mathcal{H}\Psi'+ \frac{1}{2}\Bigl( \Bar{\chi}'\delta\chi'+W^2\Bar{\mfa}'\delta\mfa'\Bigr) +\frac{1}{2} W\, W,_\chi \Bar{\mfa}'^2\delta\chi \nonumber
\\
&\qquad \qquad +\frac{a^2}{2\MPL ^2}\Bigl( 2\Phi V + V,_{\bar{\chi}}\delta\chi + V,_{\Bar{\mfa}}\delta\mfa\Bigr) =-\frac{a^2}{2\MPL ^2} \left(\delta\rho+2\Phi\bar{\rho}\right) \, , 
\\
\label{perturbed 0-i}
& k^2(\Psi'+\mathcal{H}\Phi)-\frac{k^2}{2}\Bigl( \Bar{\chi}'\delta\chi+W^2\Bar{\mfa}'\delta\mfa\Bigr)=\frac{a^2\Bar{\rho}}{2\MPL ^2}\Theta \, ,
\\
\label{perturbed i-j}
&\Psi''+\mathcal{H}\left(\Phi'+2\Psi'\right)+\left(2\mathcal{H}'+\mathcal{H}^2\right)\Phi+ \Bigl( \Bar{\chi}'^2 + W^2\Bar{\mfa}'^2 \Bigr) \Phi \nonumber
\\
&\qquad\qquad -\frac{1}{2} \Bigl( W^2\Bar{\mfa}'\delta\mfa' + \Bar{\chi}'\delta\chi' + W,_\chi W\Bar{\mfa}'^2 \delta\chi\Bigr)  
+\frac{a^2}{2\MPL ^2} \Bigl( V,_{\Bar{\chi}} \delta\chi + V,_\mfa\delta \mfa\Bigr) = 0 \,, 
\end{align}
where $W = W(\bar\chi)$ and $V = V(\bar \chi, \bar \mfa)$ and their derivatives are evaluated at the background configuration. The quantity $\Theta$ appearing in \pref{perturbed 0-i} is a sum over fluid components, $\Theta = \sum_f \Theta_f$, where $\Theta_f$ is defined for each component by $\Theta_f = -k^2v_f$ where $v_f$ is defined in \pref{4VelocityExpansion}. 


The perturbed scalar field equations similarly become
\begin{align}
 \label{perturbed dilaton}
& \delta\chi'' + 2\mathcal{H}\delta\chi' + \left[k^2-\Bar{\mfa}'^2 \left(W,_\chi ^2 + WW,_{\chi\chi}\right) +\frac{a^2}{\MPL ^2} V,_{\Bar{\chi} \Bar{\chi}}\right] \delta\chi-\Bar{\chi}' \left(\Phi'+3\Psi'\right)
 \nonumber
\\
& \qquad\qquad -2WW,_\chi \Bar{\mfa}' \delta\mfa' +\frac{a^2}{\MPL ^2} \Bigl( 2V,_{\chi}\Phi +V,_{\Bar{\chi} \Bar{\mfa}} \delta\mfa\Bigr) \\
& \qquad\qquad \qquad = -\frac{a^2}{\MPL ^2} \Bigl[\mathbf{g} _\ssB  \Bigl( \delta_\ssB +2\Phi \Bigr) \Bar{\rho}_\ssB  + \mathbf{g}_\CDM \Bigl( \delta_\CDM +2\Phi \Bigr) \Bar{\rho}_\CDM \Bigr] \, ,  \nonumber
\end{align}
and
\begin{align}
\label{perturbed axion}
&\delta\mathbf{a}''+\delta\mfa'\left(2\mathcal{H}+2\Bar{\chi}'\frac{W,_\chi}{W}\right)+\left(k^2+\frac{a^2}{\MPL ^2}\frac{V,_{\mfa\mfa}}{W^2}\right)\delta \mfa + \frac{2a^2}{\MPL ^2}\frac{V,_a\Phi}{W^2}+2\Bar{\mfa}'\frac{W,_\chi}{W}\delta\chi' \nonumber
\\
&\qquad + \delta\chi \left[2\Bar{\chi}' \Bar{\mfa}'\left(\frac{W,_{\chi\chi}}{W}-\left(\frac{W,_{\chi}}{W}\right)^2\right)+\frac{a^2}{\MPL ^2}\left(\frac{V,_{\Bar{\mfa}\Bar{\chi}}}{W^2}-\frac{W,_\chi}{W^3}V,_{\Bar{\mfa}}\right)\right]
-\left(\Phi' + 3\Psi'\right) \Bar{\mfa}' \nonumber
\\
& \qquad \qquad =-\frac{a^2}{\MPL ^2 W^2} \Bar{\rho}_\ssB \left[\mathbf{g_\mfa}\left(\delta_B+2\Phi-2\frac{W,_\chi}{W} \delta\chi  + \frac{\mathbf{g_\mfa}_{,\mfa} \delta\mfa}{\mathbf{g_\mfa}} \right)\right] \, .
\end{align}
Finally, the perturbed continuity equations for baryons and cold dark matter are
\begin{align}
\label{Continuity equation perturbed B}
 & \delta_B'+\Theta_B-3\Psi'  = \mathbf{g} _\ssB \delta\chi'+\mathbf{g_\mfa}\delta\mfa' + \bar{\mfa} '  \mathbf{g_\mfa}_{,\mfa} \delta\mfa \, , 
\\
\label{Continuity equation perturbed CDM}
& \delta_\CDM' + \Theta_\CDM -3\Psi'  = \mathbf{g}_\CDM\delta\chi' \, ,
\end{align}
while the Euler equations for these fluid components are
\begin{align}
\label{Euler equation Baryon}
& \Theta'_B+\Theta_B\mathcal{H}-k^2\Phi=- \Bigl[\mathbf{g} _\ssB \left(\Bar{\chi}'\Theta_B-k^2\delta\chi\right)+\mathbf{g_\mfa}\left(\Bar{\mfa}'\Theta_B-k^2\delta\mfa\right)\Bigr] + J_{(\ssB){\rm eq}}\, , 
\\
& \Theta'_\CDM+\Theta_\CDM\mathcal{H}-k^2\Phi=-\mathbf{g}_\CDM \left(\Bar{\chi}'\Theta_\CDM-k^2\delta\chi\right) \,, \label{Euler equation CDM}
\end{align}
where \cite{Ma:1995ey}
\begin{equation}
   J_{(\ssB){\rm eq}} = \frac{4 \bar \rho_\gamma}{3 \bar \rho_\ssB} \, a \, n_e \, \sigma_\ssT \Bigl( \Theta_\gamma - \Theta_\ssB \Bigr) \,,
\end{equation}
and $\sigma_\ssT$ is the Thomson cross section (including the field-dependence of the electron mass). The evolution of $\Theta_\gamma$ does not depend at all on the scalar fields because the conformal invariance of the Maxwell action implies photons do not directly couple to the dilaton (and we assume no axion-photon coupling), and because the much larger entropy of the photon fluid means it is largely unchanged by the energy exchange with the baryons. 

These are the equations we code into CLASS \cite{Diego_Blas_2011} when seeking implications of these models for the CMB. We do not discuss radiation or neutrino components to the cosmic fluid in the above because for these we allow CLASS to evolve perturbations without new scalar-field complications.\footnote{Leaving neutrino evolution unchanged in axiodilaton models is a simplifying assumption and is not required. Indeed, reasonable choices for neutrino-scalar interactions in these models allow -- but do not require -- neutrino masses to be of the same order of magnitude as the dark energy density. We nonetheless defer studies of changes to cosmological neutrino evolution for future work. } 

\subsection{The quasistatic r\'egime}

Before turning to the numerical evolution of the equations described above, it is helpful to examine how they appear under the \emph{quasistatic} (QS) approximation \cite{Noller:2013wca},  which is particularly relevant for the later stages of the cosmological evolution such as large-scale structure formation.

The QS approximation applies deep inside the cosmological horizon, $k^2\gg a^2H^2$, which is the regime most relevant to the formation of large-scale structure. The observation is that in this regime the time derivatives of fields evolving on Hubble times (such as the gravitational potential) can be dropped relative to spatial gradients. More specifically, the two assumptions underlying the QS approximation are
\begin{equation}
    |X'|\lsim \mathcal{H}\left|X\right| \qquad \hbox{and} 
    \qquad 
    k^2\gg\mathcal{H}^2 \,,
\end{equation}
where $X$ represents any field in the sub-horizon limit, which are considered to be sufficiently slow-varying and lack highly oscillatory behaviour.

Since structure formation is of interest, we assume in this section that the energy density, $\bar\rho_m = \bar\rho_\ssB + \bar\rho_{\CDM}$, in nonrelativistic matter is larger than that in radiation. In this regime $\bar \rho \simeq \bar \rho_m$ and the conservation equations like \pref{contequ CDM} and \pref{contequ B} imply that the fluid densities evolve adiabatically and so remain approximately constant over sub-Hubble timescales. Neglect of time derivatives in the Friedmann and background scalar equations then implies the values of the scalar fields and of the Hubble scale are set by
\begin{equation}
\label{qsa background}
 \mathcal{H}^2 \simeq \frac{ a^2}{3 \MPL^2 } \Bigl( \bar \rho_m +  V \Bigr)  \,,
 \quad
V,_{{\chi}} \, \simeq -\textbf{g}\bar{\rho}_\ssB -\textbf{g}_\CDM \bar{\rho}_\CDM \, ,  
\quad
 V,_\mfa \,\simeq -\textbf{g}_{\bar{\textbf{a}}}\bar{\rho}_\ssB \,.
\end{equation}
%
%
In terms of these background values we define the quantities 
\begin{equation}
m_\chi^2 = \frac{V,_{\chi\chi}}{\MPL^2} \, , \quad m^2_{\chi\mfa} = \frac{V,_{\chi \mfa}}{\MPL^2} \qq{and}  m^2_{\mfa}  =  \frac{V,_{\mfa\mfa}}{\MPL^2} \, ,
\end{equation}
and also define
\begin{equation}
  k_\mfa^2 \equiv \frac{k^2}{a^2}+\frac{m_{\mfa}^2}{W^2} + \frac{\mathbf{g}_{\mfa,\mfa}\Bar{\rho}_\B}{W^2} \qq{and}  k_\chi^2 \equiv \frac{k^2}{a^2}+m_\chi^2 \,.
\end{equation}
With these definitions eqs.~(\ref{Perturbed Friedmann}), \pref{perturbed dilaton} and \pref{perturbed axion} become the quasi-static versions of the perturbed Poisson, dilaton and axion equations, respectively: 
\begin{align}
\label{qsa friedmann}
 k^2\Psi & \approx -\frac{a^2}{2 \MPL ^2}\Bigl[ \bar{\rho}_\CDM \delta_\CDM + \bar{\rho}_B\delta_\ssB - \Bigl(\textbf{g}_\CDM \bar \rho_\CDM + \textbf{g} _\ssB \bar{\rho}_B\Bigr) \delta\chi-\textbf{g}_\mfa\bar{\rho}_B\delta \mfa\Bigr] \, , 
    \\
\label{qsa dilaton}
\delta\chi & \approx-\frac{1}{k_\chi^2} \left[\frac{1}{\MPL ^2} \Bigl( \textbf{g}_\CDM \bar \rho_\CDM \delta_\CDM + \textbf{g} _\ssB  \bar \rho_\ssB \delta_\ssB \Bigr)  + m_{\chi\mfa}^2 \delta\mfa \right] \, , 
\\
\label{qsa axion}
\delta\mfa & \approx -\frac{1}{W^2 k_\mfa^2}\left(m_{\chi\mfa}^2\delta\chi+\textbf{g}_\mfa\frac{\bar{\rho}_B}{\MPL ^2}\delta_B-\textbf{g}_\mfa\frac{W,_\chi}{W}\frac{\bar{\rho}_B}{\MPL ^2}\delta\chi\right) \,. 
\end{align}
Because we work in the quasistatic approximation we have ignored terms that vary on the Hubble time scale, such as $({\chi'}/{a^3})\left[ a^3 \delta \chi\right]'$ in the Poisson equation \cite{Noller:2013wca}. 

Finally, the Newtonian limits of (\ref{Continuity equation perturbed B}) and (\ref{Continuity equation perturbed CDM}) are equivalent to dropping time derivatives in the conservation equations for each species, resulting in
\begin{equation}\label{qsa first fluid}
    \delta'_\ssB \simeq -\Theta_\ssB \qq{and} 
    \delta'_\CDM \simeq -\Theta_\CDM \,.
\end{equation}
Substituting these results into the relevant Euler equation (\ref{Euler equation Baryon}) and (\ref{Euler equation CDM}), in which we drop no terms, gives the quasistatic version of the growth equations for $\delta_B$ and $\delta_\CDM$. Explicitly, using (\ref{qsa friedmann}), (\ref{qsa dilaton}) and (\ref{qsa axion}) gives
\begin{eqnarray}
 \label{Delta B general}
&&\delta_\ssB'' + \delta_\ssB' \left[\mathcal{H} +\mathbf{g} _\ssB \Bar{\chi}' + \mathbf{g_a} \Bar{\mfa}'\right] = 4\pi a^2 G_{\rm eff(\ssB)}^B\Bar{\rho}_\ssB\delta_\ssB+4\pi a^2G_{\rm eff(\ssB)}^\ssC\Bar{\rho}_\ssC\delta _\ssC   \, ,
\\
\label{delta C general}
&&\delta_\ssC''+\delta_\ssC'\left[\mathcal{H}+\mathbf{g}_\CDM \Bar{\chi}'\right] = 4\pi a^2 G_{\rm eff (\ssC)}^\ssB\Bar{\rho}_\ssB\delta_\ssB+4\pi a^2 G_{\rm eff (\ssC)}^\ssC\Bar{\rho}_\ssC\delta _\ssC \, ,
\end{eqnarray}
where the effective gravitational constants relevant for the evolution of $\delta_\ssB$ are 

\begin{eqnarray}
\label{GeffBB}
  G_{\rm eff(\ssB)}^\ssB &=& \frac{1}{8 \, \pi  \MPL ^2}   \bigg\{1- \frac{1}{\mathcal{A}^2}\bigg[\mathbf{g} _\ssB  \left(\frac{2 \, k^2}{a^2}+\frac{\mathbf{g}_\CDM}{\mathbf{g} _\ssB }\frac{\Bar{\rho}_\ssC}{ \MPL ^2}+\frac{\Bar{\rho}_\ssB}{ \MPL ^2}\right) \left(\frac{\mathbf{g_\mfa}}{W^2} m_{\chi\mfa}^2 - \mathbf{g} _\ssB  k_\mfa^2 \right) \nonumber\\
  &&\qquad + \mathbf{g_\mfa} \left(\frac{2 \, k^2}{a^2}+\frac{\Bar{\rho}_\ssB}{\MPL ^2}\right) 
 \left(\frac{\mathbf{g} _\ssB }{W^2}\left(m_{\chi\mfa}^2-\frac{W,_\chi}{W}\mathbf{g_\mfa}\frac{\Bar{\rho}_\ssB}{\MPL ^2}\right)-\frac{\mathbf{g_\mfa}}{W^2}k_\chi^2\right)\bigg] \bigg\} \, ,
 \\
 \label{GeffBC}
 G_{\rm eff(\ssB)}^\ssC &=&\frac{1}{8\pi  \MPL ^2} \bigg\{1- \frac{1}{\mathcal{A}^2}  \bigg[- \mathbf{g} _\ssB \mathbf{g}_\CDM k_\mfa^2  \left(\frac{2\, k^2}{a^2}+\frac{\mathbf{g}_\CDM}{\mathbf{g} _\ssB }\frac{\Bar{\rho}_\ssC}{ \MPL ^2}+\frac{\Bar{\rho}_\ssB}{\MPL ^2}\right)  \nonumber \\
 &&\qquad + \frac{ \mathbf{g_c} \, 
 \mathbf{g_\mfa}}{W^2}\left(\frac{2 \, k^2}{a^2}+\frac{\Bar{\rho}_\ssB}{\MPL ^2}\right) 
 \left(m_{\chi\mfa}^2- \mathbf{g_\mfa} \frac{W,_\chi}{W}\frac{\Bar{\rho}_\ssB}{\MPL ^2}\right) \bigg] \bigg\} \,  ,
\end{eqnarray}
while those appearing in the evolution of $\delta_\CDM$ are 
\begin{eqnarray}
\label{GeffCB}
G_{\rm eff(\ssC)}^\ssB & = &\frac{1}{8\pi \MPL ^2}\bigg\{1- \frac{1}{\mathcal{A}^2}\bigg[ \mathbf{g}_\CDM \left(\frac{2 \, k^2}{a^2} + \frac{\Bar{\rho}_\ssC}{ \MPL ^2}+ \frac{\mathbf{g} _\ssB }{\mathbf{g}_\CDM}\frac{\Bar{\rho}_\ssB}{ \MPL ^2}\right)\left(\frac{\mathbf{g_\mfa}}{W^2} m_{\chi\mfa}^2 - \mathbf{g} _\ssB  k_\mfa^2\right)  \nonumber
\\
&&\qquad + \frac{\Bar{\rho}_\ssB}{\MPL ^2}\mathbf{g_\mfa}\left(\frac{\mathbf{g}_\ssB}{W^2}\left(m_{\chi\mfa}^2-\frac{W,_\chi}{W}\mathbf{g_\mfa}\frac{\Bar{\rho}_\ssB}{\MPL ^2}\right)-\frac{\mathbf{g_\mfa}}{W^2}k_\chi^2\right)\bigg]\bigg\} \, ,
\\
\label{GeffCC}
G_{\rm eff(\ssC)}^\ssC & = &\frac{1}{8\pi \MPL ^2}\bigg\{1-\frac{1}{\mathcal{A}^2} \bigg[ -\mathbf{g}_\CDM^2  k_\mfa^2 \left(\frac{2 \, k^2}{a^2} \nonumber
+ \frac{\Bar{\rho}_\ssC}{ \MPL ^2}+ \frac{\mathbf{g} _\ssB }{\mathbf{g}_\CDM}\frac{\Bar{\rho}_\ssB}{ \MPL ^2}\right)
\\
&&\qquad + \mathbf{g}_\CDM \frac{\Bar{\rho}_\ssB}{\MPL ^2}\frac{\mathbf{g_\mfa}}{W^2}\left(m_{\chi\mfa}^2- \mathbf{g_\mfa} \frac{W,_\chi}{W}\frac{\Bar{\rho}_\ssB}{\MPL ^2}\right)\bigg]\bigg\} \,.
\end{eqnarray}
These equations define the quantity
\begin{equation}
\mathcal{A}^2 =k_\mfa^2 k_\chi^2-\left(\frac{m_{\chi\mfa}^2}{W^2}-\frac{W,_\chi}{W}\frac{\mathbf{g_\mfa}}{W^2}\frac{\bar{\rho}_\ssB}{\MPL ^2}\right)m_{\chi\mfa}^2   \, .  \end{equation}
In eqs.~\pref{Delta B general} and \pref{delta C general} the first-derivative terms on the left-hand side show that the friction experienced by the fluctuations is modified due to the couplings of the two fluids to the scalar fields. This leads to the matter species experiencing a cosmological `drag' that influences the typical Newtonian motion of the particles.  

These expressions also reveal how the scalar fields modify the effective gravitational Newton's constant that controls the strength of the attraction towards local overdensities. These modifications, given in eqs.~\pref{GeffBB} through \pref{GeffCC}, arise due to the additional scalar-moderated forces acting between matter species. The scalar field responsible for the modification can be identified by the coupling strength appearing in each term. For instance, terms proportional to $\mathbf{g}^2$ and $\mathbf{g}_\mfa^2$ describe the effects of direct scalar-matter couplings. Terms proportional to $\mathbf{g \, g_\mfa}$ indicate indirect interactions allowed by mixing between the axion and dilaton, such as is produced by nonzero cross derivatives like $V_{,\mfa\chi}$. Terms cubic in coupling constants also have their roots in indirect interactions, this time mediated by nonzero $W_{,\chi}$ in the axio-dilaton kinetic coupling.

\section{ Evolution with constant axion--baryon coup\-lings}
\label{Section:Yoga Case}

Since the CMB is the grave on which most models come to die, a serious study of axio-dilaton cosmologies must examine both the background and the evolution of cosmological perturbations. This section describes the results of such an analysis, performed using the Cosmic Linear Anisotropy Solving System code, CLASS \cite{Diego_Blas_2011}, modified to compute the dynamics of multiple scalar fields and their couplings to matter. 


Recall that in addition to the changes to the perturbation equations discussed in the previous section, our analysis also takes into account the evolution of the electron mass. A time-varying electron mass affects the recombination history by changing the recombination redshift\footnote{Such changes needn't be bad; \cite{Schoneberg:2021qvd} uses them to help with the Hubble tension.}  $z_*$. In this work we do not perform a comprehensive data analysis using Planck and other cosmological data. Instead we focus on identifying axio-dilaton models that show promise and highlight their features for future work. 

\subsection{Benchmark models}

Up to this point we have kept our options open for the precise form taken by the functions $W(\chi)$, $V(\chi,\mfa)$ and $\cU(\mfa)$ and on the size and species dependence of the dilaton-matter couplings $\mathbf{g}_B$ and $\mathbf{g}_\CDM$. We now focus on a more specific class of four benchmark models in order to explore their predictions in more detail. To this end we zero in on the class of Angle-Saxion models discussed in section \ref{yoga models}, for which $W(\chi)=e^{-\zeta\chi}$ and $\zeta$ is a parameter to be specified. This includes in particular the Yoga models \cite{Burgess:2021obw} for which $\zeta$ is predicted to be $\zeta = \sqrt{2/3}$.

To further specify the model we choose the vacuum scalar potential. In this section we assume $V$ to be independent of the axion -- but we revisit this choice and introduce an axion potential in \S\ref{QuadraticAxionCouplingSection} below. We further assume the dilaton dependence to have the form
\begin{equation} \label{DilatonPot}
V=Ue^{-\lambda\chi} \, ,
\end{equation}
with $\lambda = 4\zeta$ and the prefactor $U$ chosen to be a quadratic polynomial of $\chi$ 
\begin{equation} \label{DilatonPotPrefactor}
U(\chi) = V_0\left[1-u_1\chi+\frac{u_2}{2}\chi^2\right] \,,
\end{equation}
along the lines used in \cite{Albrecht:1999rm, Skordis:2000dz, Albrecht_2002, Gasperini:2001pc} (and more recently\footnote{Refs.~\cite{Albrecht:1999rm, Skordis:2000dz} correspond to the special case of \cite{Albrecht_2002} evaluated at $\bfg = 0$ (no matter-dilaton coupling), and all three of these differ from \cite{Burgess:2021obw} by omitting the axion and in the size taken for $\lambda$, with the earlier papers choosing $\lambda \sim \cO(0.01)$ and \cite{Burgess:2021obw} taking $\lambda \sim \cO(1)$. } in \cite{Burgess:2021obw}). With these papers in mind we choose coefficients to ensure a local minimum that gives the present-day dark energy density, doing so for field values $\chi_{min} \sim 70$. (Although not required by cosmology such a large size of $\chi_{min}$ is used in these models to provide a common explanation for the size of the electroweak, neutrino and dark energy hierarchies \cite{Burgess:2021obw}.) 

Without the prefactor $U$ the dilaton would continuously roll down the exponential slope in the absence of the other parts of the cosmological fluid. This need not be a problem if $\lambda$ is small enough because slow-roll evolution down an exponential potential produces an equation of state parameter $w+1 \simeq \frac13 \lambda^2/(1+\frac16 \lambda^2)$. Requiring $w+1 \lsim 0.1$ implies $\lambda  \lsim 0.5$. So constant $U$ is not viable for the Yoga-model choices $\lambda = 4\zeta$ and $\zeta = \sqrt{\frac23} \simeq 0.8$, but it potentially becomes an option if $\lambda$ is not too much smaller -- a case considered in detail in \S\ref{weakly coupled axiodilatons section}. For larger $\lambda$ a minimum for $U$ is required to allow the potential to dominate the energy density at late times. In what follows we work with \pref{DilatonPotPrefactor} so we can explore cosmologies with both small and large $\lambda$.

Our final choice for this section is to choose the form of the axion-matter coupling. Although $\mathbf{g_a}$ as defined above is typically a periodic function of $\mfa$, it can suffice to approximate it as a constant if $\mfa$ does not move very far during the cosmological epochs of interest. It could be chosen to be an approximately linear function of $\mfa$ if it does not move far but starts near a zero of $\ga$. 

In this section we assume the axion is not near a zero of $\ga$ and so approximate $\ga$ as being constant. (This choice is relaxed in \S\ref{QuadraticAxionCouplingSection}, which explores the vicinity of a zero of $\ga$.) More specfically, we choose the function $\cU = (1+\gamma \,\mfa)$ and choose $\gamma$ small enough that $\gamma \mfa$ remains very small throughout the observable history of the universe. We check {\it ex post facto} that $\gamma \mfa$ actually remains small. With this choice baryon masses and baryon-axion couplings are given by 
\begin{equation} \label{eq:linga}
  m_{\ssB} = m(1+\gamma \, \mfa) \, e^{\mathbf{g} _\ssB \chi} \simeq m \, e^{\mathbf{g} _\ssB \chi} 
  \quad \hbox{and} \quad
    \mathbf{g}_\mfa = \frac{\cU'}{\cU} = \frac{\gamma}{1+\gamma \,\mfa} \simeq \gamma .
\end{equation}
Requiring $\gamma \mfa$ to be small also ensures the cosmological evolution of baryon masses is dominated by the evolution of the dilaton $\chi$. In all of our benchmark models we choose $\lambda = 4 \zeta$ so the free parameters are\footnote{We note in passing that in our conventions the choice $W_{\rm min} = W(\chi_{\rm min}) \sim (\MEW/\MPL)^2 \sim 10^{-28}$ implies the kinetic-term decay constant is $f = M_p W_{\rm min} \sim 0.1$ eV and so $\gamma \sim 10^{-28}$ corresponds to the decay constant relevant to present-day matter interactions being $F = f/\gamma \sim M_p$.} the values of $\bfg _\ssB $, $\bfg_\CDM$, $\gamma$ and $\zeta$. 

In the remainder of this section we explore the cosmological implications of these choices, focussing on the following four specific cases (see also Table \ref{models}):
\begin{enumerate}
    \item {\bf Universal Yoga Model:} This model is defined by the choices $\bfg:= \bfg_\ssB = \bfg_\ssC = - \frac12 \, \zeta$ and $\zeta = \sqrt{\frac23}$. We choose the axion-matter coupling to be $\gamma = - 8 \times 10^{-29}$.
    \item {\bf Opposite-coupling Yoga Model:} This model is defined by the choices $\bfg_\ssB = - \frac12 \, \zeta$ and $\zeta = \sqrt{\frac23}$ but with $\bfg_\CDM = - \bfg_\ssB/x$ with $x$ of order 1-10. The axion-matter coupling is again $\gamma = - 8 \times 10^{-29}$.
    \item  {\bf Small-$\zeta$ Angle-Saxion Model:} This model is defined by the choices $\bfg := \bfg_\ssB = \bfg_\CDM = - \frac12 \, \zeta$ but where $\zeta = x$ is varied through values much smaller than unity. Reducing $\zeta$ causes a large increase in the axio-dilaton kinetic coupling $W(\chi)$, increasing the decay constant $F$. To compensate for this we take  $\gamma = -4\times10^{-2}$ in this case.
    \item  {\bf General coupling Angle-Saxion Model:} This model is defined by the choices $\bfg:= \bfg_\ssB = \bfg_\CDM = x$ and $\zeta = \sqrt{\frac23}$ where $x$ is varied through values much smaller than unity. The axion-matter coupling is again $\gamma = - 8 \times 10^{-29}$.
\end{enumerate}

For each of these choices we first numerically integrate the background configurations to verify that a viable background cosmology is possible for which the background densities and equations of states (in particular for the Dark Energy) are acceptable. We require the background parameters to reproduce the present-day values $H_0 = 100h\; \rm km/s\;\hbox{Mpc}^{-1}$ with $h = 0.6756$, $\Omega_\ssB h^2 = 0.022$, and $\Omega _\ssC h^2 = 0.12$. For large $\lambda = 4\zeta$ and/or $\mfg_\ssB$ the scalar potential and/or matter couplings tend to strongly drive the dilaton to larger values so the issue is whether or not there are initial conditions for which the dilaton gets trapped at the minimum of its potential and so achieves the correct late-time equation of state for Dark Energy.  

For those models with successful background evolutions we then numerically evolve the cosmological fluctuations, assuming adiabatic gaussian initial conditions for the power spectrum taken from the 2018 Planck LCDM best fit \cite{2020}. This sets the spectral index $n_s = 0.966$, the pivot scale $k_{piv} = 0.05\;\rm M_{pc}^{-1}$ and scalar amplitude $A_s = 2.10\times10^{-9}$.

The results of these simulations are now described in detail, fleshing out the brief summary given in Table \ref{models}.

\subsection{Background evolution}

The background axio-dilaton equations of motion with the above choices are
\begin{equation}
\label{friedmann dilaton eval}
 \Bar{\chi}''+2\mathcal{H}\Bar{\chi}'+\zeta e^{-2\zeta\chi}\Bar{\mfa}'^2+\frac{a^2}{\MPL ^2}\Bigl( V,_\mathbf{\chi} +\mathbf{g}_\CDM \, \bar\rho_\CDM +\mathbf{g} _\ssB  \, \bar\rho_\ssB \Bigr) =0 \, ,   
\end{equation}
with $V$ given by eqs.~\pref{DilatonPot} and \pref{DilatonPotPrefactor} and
\begin{equation}
\label{friedmann axion eval}
\Bar{\mfa}''+2\mathcal{H}\Bar{\mfa}'-2\zeta\Bar{\mfa}'\Bar{\chi}'+ \mathbf{g_\mfa} \, \Bar{\rho}_\ssB \, \frac{a^2}{\MPL ^2 } \, e^{2\zeta\chi} =0. 
\end{equation}
We know that viable background cosmologies for these equations should exist for small enough $\zeta$ and $\bfg$ because the implications of dilaton evolution \pref{friedmann dilaton eval} (with matter couplings but without an axion) were studied in some detail some time ago \cite{Albrecht_2002}. Viable cosmologies were possible provided $\zeta \lsim \mathcal{O}(0.01)$. In this limit both the scalar potential and the matter coupling tend to drive $\chi$ to larger values but for small $\zeta$ Hubble friction can prevent the dilaton from over-shooting the minimum, allowing it to be trapped and to become a late-time cosmological constant. Dilaton overshoot could not be avoided when $\zeta \sim \cO(1)$.

Ref.~\cite{Burgess:2021obw} explored background cosmologies including also the axion, assuming slightly different couplings than we use here (\cite{Burgess:2021obw} chose an axion-matter source ${\cal J} = \bfg_y n_B$ for constant $\bfg_y$, without specifying that this coupling arose from an axion-dependent matter mass). They found that the presence of the axion allowed viable cosmologies to become possible even for $\zeta \sim \cO(1)$. The axion makes all the difference because the presence of $W(\chi)$ implies a nonzero $\bar\mfa'$ introduces a new effective potential for the evolution of $\chi$ and unlike all the others this pushes $\chi$ to smaller values. It is the existence of this resistance to $\chi$ evolving to larger values that allows the dilaton not to overshoot despite the stronger force from the potential that larger $\zeta$ implies. If the coupling chosen in \cite{Burgess:2021obw} is interpreted as arising due to an axion-dependent mass it corresponds in the present notation to $\mathbf{g_a} = \bfg_y/ m_\ssB(\mfa)$. We do not here find the same successful cosmologies found in \cite{Burgess:2021obw} for the reasons outlined in the next subsection.

\subsubsection{Yoga models: universal couplings}

We are unable to find a viable background cosmology for the Universal Yoga model (Model 1 of Table \ref{models}). The cosmology fails for one of two complementary reasons: either the axion evolution unacceptably drains away the baryon density or the dilaton never gets trapped at the potential's minimum and so the dilaton energy density never resembles the Dark Energy. 

In the absence of the dilaton-axion coupling the dilaton runaway arises -- despite not doing so for \cite{Albrecht:1999rm, Skordis:2000dz, Albrecht_2002} -- because the choice $\zeta = \sqrt{\frac23}$ is much larger than the values for $\zeta$ entertained in these earlier papers. The larger value of $\zeta$ causes a dilaton runaway because in the early universe the dilaton-matter coupling dominates over the vacuum potential and it pushes $\chi$ to large values so quickly that it is unable to be trapped by the local minimum of the scalar potential by the time the matter density is small enough to allow the potential eventually to compete.

\begin{figure}[hbt!]
    \centering
     \includegraphics[width=\textwidth]{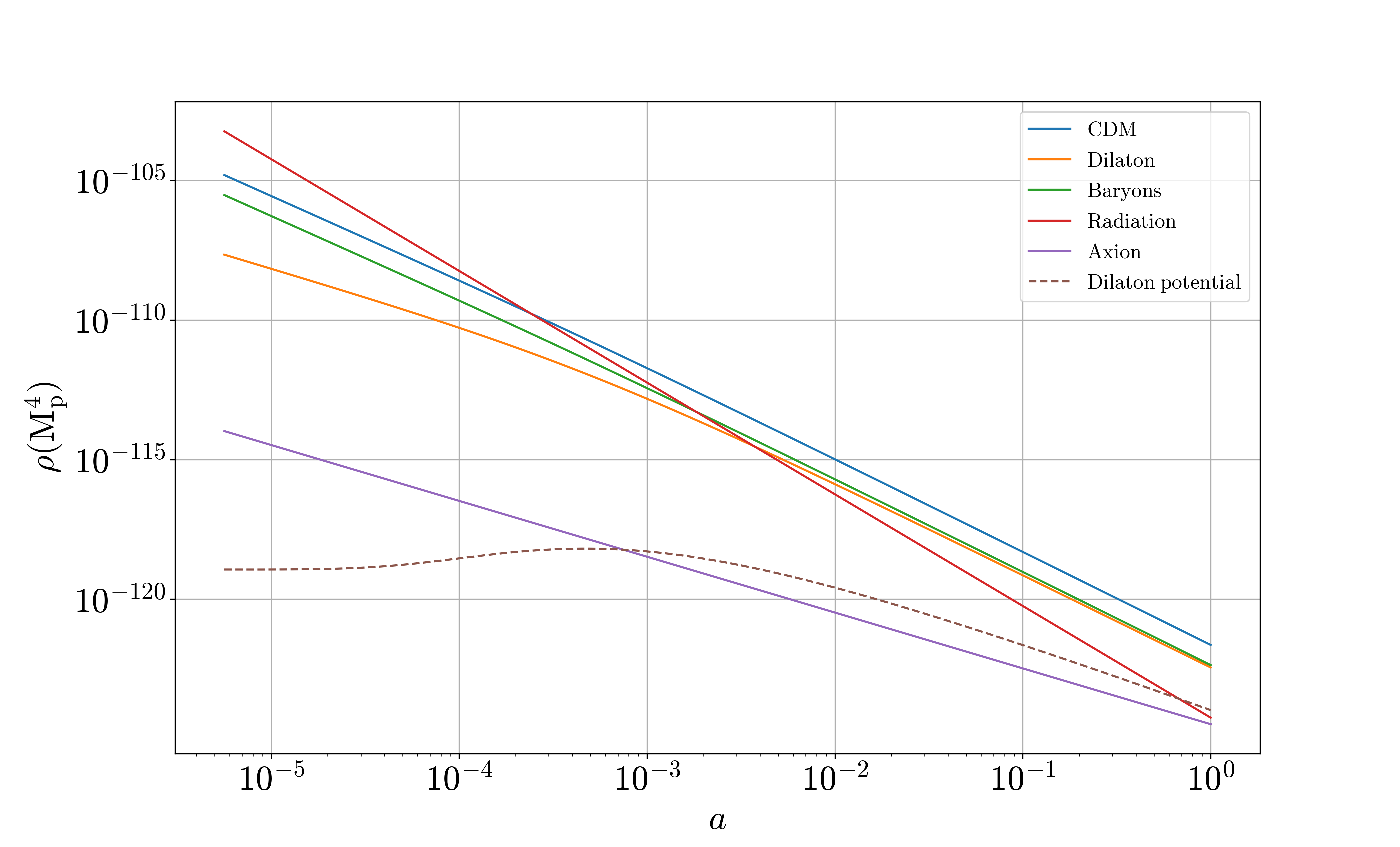}
     \caption{\small Background energy density evolution in the Universal Yoga model in baryons (green), CDM (blue), radiation (red), dilaton's total (orange), dilaton's potential (dashed), and axion (purple)  with $\mathbf{g}\equiv\mathbf{g}_\ssB = \mathbf{g}_\CDM = -\sqrt{1/6}$, $\zeta = \sqrt{2/3}$ and $\gamma =-10^{-29}$.}
     \label{fig:dilaton pulled out of well}

\end{figure}

As noted above, the dilaton runaway can be remedied if the axion is rolling quickly enough because the axion-dilaton coupling mediated by $W$ pushes the dilaton to smaller values. And a rolling axion can be obtained by dialing up the axion-matter coupling, because the nonzero background baryon density then acts as a source for the axion, winding it up and causing it to roll in the absence of an axion potential. 

A closer examination of (\ref{friedmann axion eval}) provides further insight into how baryons wind up the axion. Before matter-radiation equality, the dominant term comes from the Hubble friction, while matter remains sub-dominant. The third term in the equation, being oscillatory, has minimal impact during this phase. However,  as we progress to late times, where $\rho_m/\MPL^2 \sim \mathcal{H}^2$, the absence of an axion potential along with the dilaton oscillating within its potential well,  ensures that the term sourcing the axion dominates in (\ref{friedmann axion eval}). In such a scenario, (\ref{friedmann axion eval}) describes the exponential growth of the axion velocity proportional to the baryon energy density owing to the additional factor $e^{2\zeta\chi}$ resulting from the modified geometry of the target field space. 

Any increase in the axion velocity comes at the expense of baryon energy, as may be seen from the baryon continuity equation in (\ref{contequ B}), causing the baryon energy density to fall much too quickly. Keeping this drainage acceptable requires dialing down $\ga$, since this reduces the baryon draining effect by limiting the axion velocity. However, this takes us back to the original problem wherein the axion rolls too slowly to prevent the dilaton runaway. Alleviating both problems requires reducing {\it both} matter-axion and matter-dilaton coupling strengths (which takes us beyond the domain of the Universal Yoga model). These contradictory demands on $\ga$ have prevented our finding viable background cosmologies for the Universal Yoga scenario, despite the potentially stabilizing effects of the axion.

There is a caveat however because when axion motion drains energy too efficiently from the baryons it does so only because the baryon mass evolves significantly as the axion rolls. After all, it is only the time-dependence of the baryon mass that allows $\rho_\ssB$ to deviate from its usual $1/a^3$ falloff as the universe expands. Furthermore, the baryon mass is unbounded from below when it is modelled by \pref{eq:linga} because it depends linearly on $\mfa$ when $\ga$ is approximately constant. Catastrophic energy loss to axions only occurs when $\gamma \mfa$ does {\it not} remain small and this means the obstruction to building successful cosmologies is being found in a regime where the assumption that $\ga$ remain constant is a poor approximation for the periodic potential $\cU(\mfa)$ Yoga-type models actually predict, even if its size is chosen small enough to evade constraints on baryonic mass variation. Further exploration of Universal Yoga cosmology using more realistic oscillatory masses is warranted to close this loophole.

\begin{figure}[hbtp]
    \centering
     \begin{subfigure}[b]{\textwidth}
         \centering
         \includegraphics[width=\textwidth]{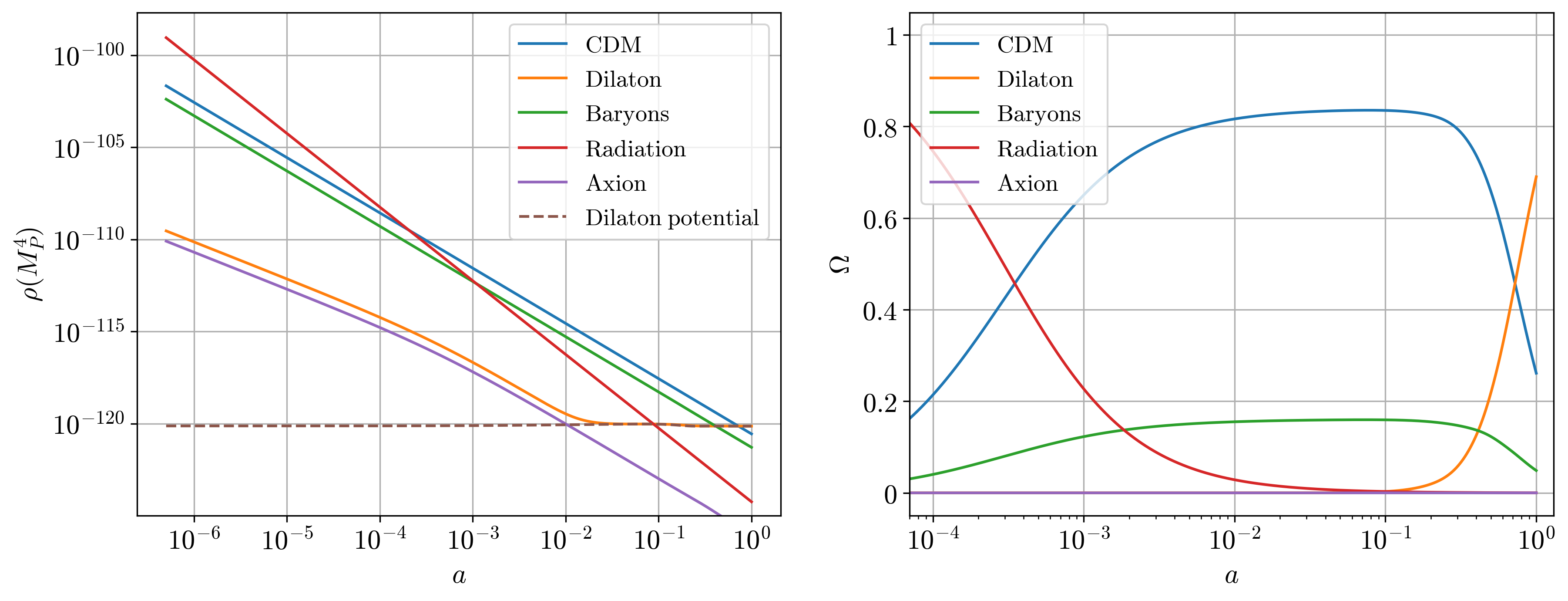}
     \end{subfigure}
     \hfill
     \centering
     \begin{subfigure}[b]{\textwidth}
         \centering
         \includegraphics[width=\textwidth]{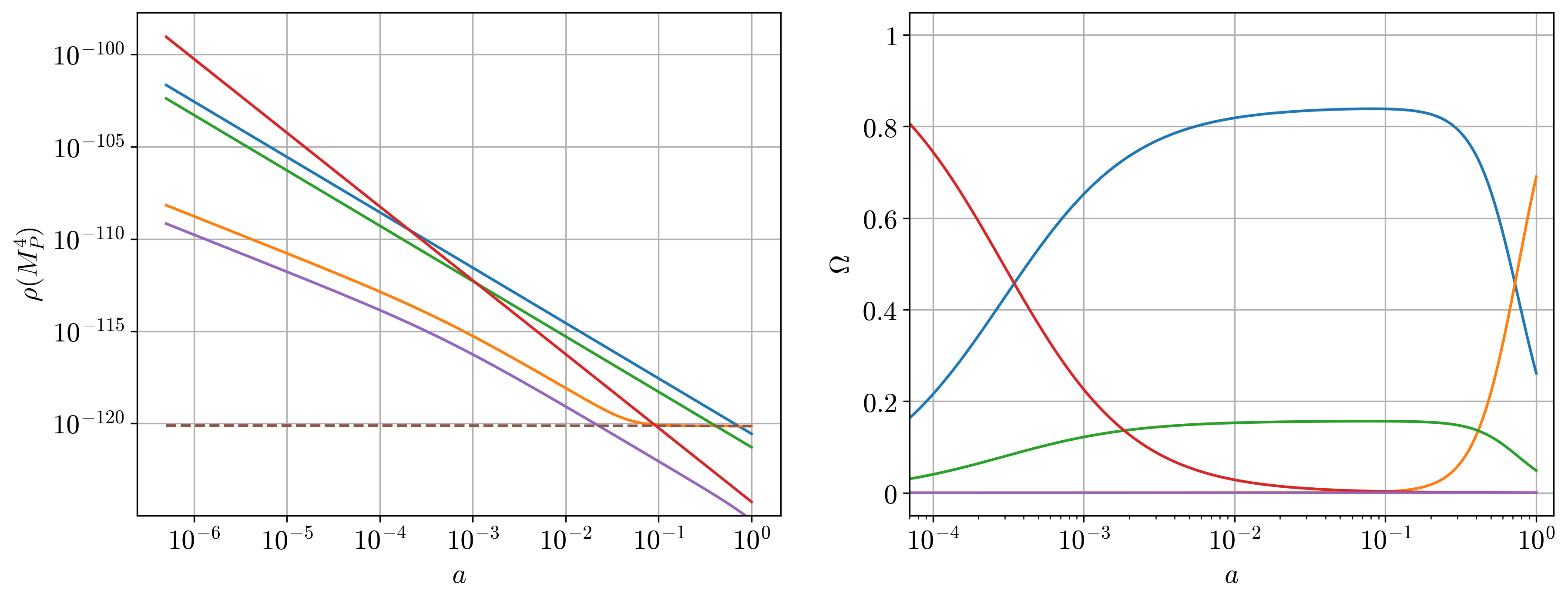}
     \end{subfigure}
     \hfill
     \centering
     \begin{subfigure}[b]{\textwidth}
         \centering
         \includegraphics[width=\textwidth]{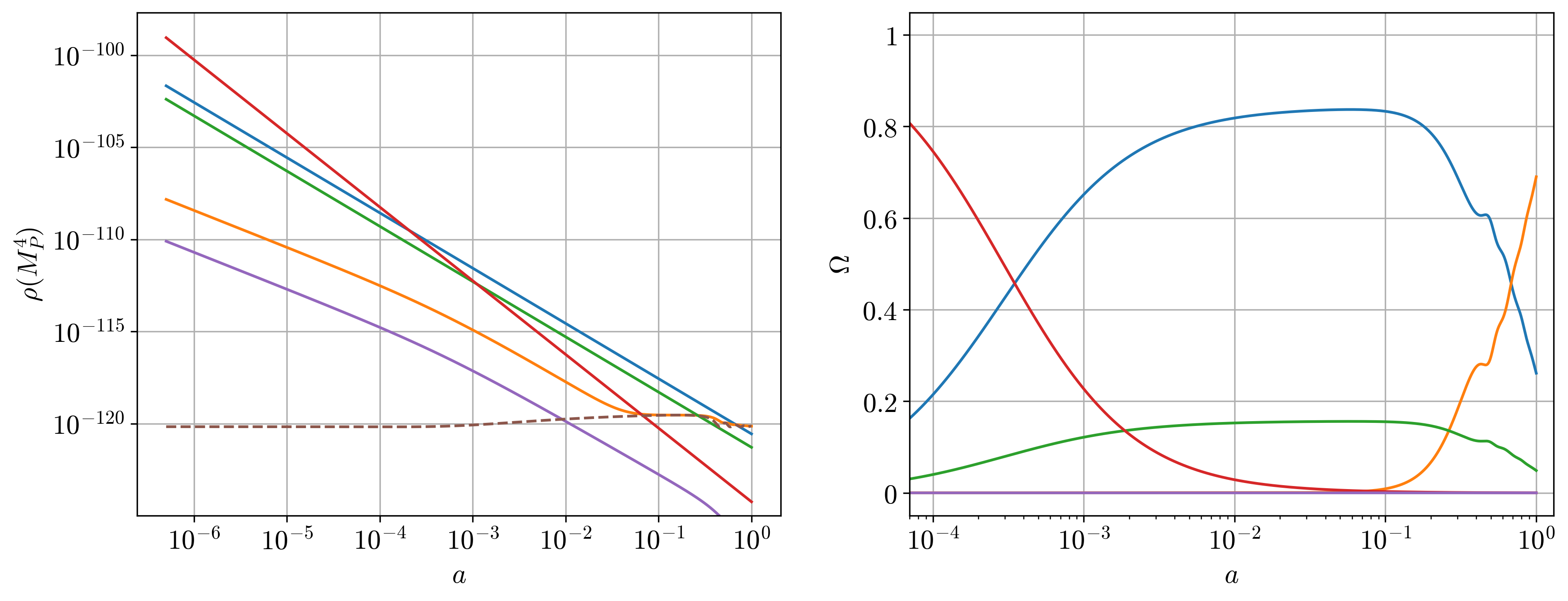}
     \end{subfigure}
    \caption{{\small Background evolution for the three viable cases in table \ref{models}. Top row: Model 2 with couplings $\mathbf{g_\CDM} = -\mathbf{g} _\ssB /5$, $\mathbf{g} _\ssB  = -\zeta/2$ and $\zeta = \sqrt{2/3}$ with $\gamma = -8\times10^{-29}$. Middle row: Model 3 with couplings $\mathbf{g} \equiv \mathbf{g} _\ssB  = \mathbf{g}_\CDM = -\zeta/2$ and $\zeta = 0.04$, with $\gamma = -4\times10^{-2}$. Bottom row: Model 4 with couplings $\mathbf{g} = -0.03$ and $\zeta = \sqrt{2/3}$, with $\gamma = -8\times10^{-29}$. All cases show the evolution of energy densities in baryons (green), CDM (blue), radiation (red), dilaton's total energy (orange), dilaton's potential energy (dashed), and axion total energy (purple).}}
        \label{fig:Backgrounds}
\end{figure}

\begin{figure}[hbt!]
    \centering
     \includegraphics[width=\textwidth]{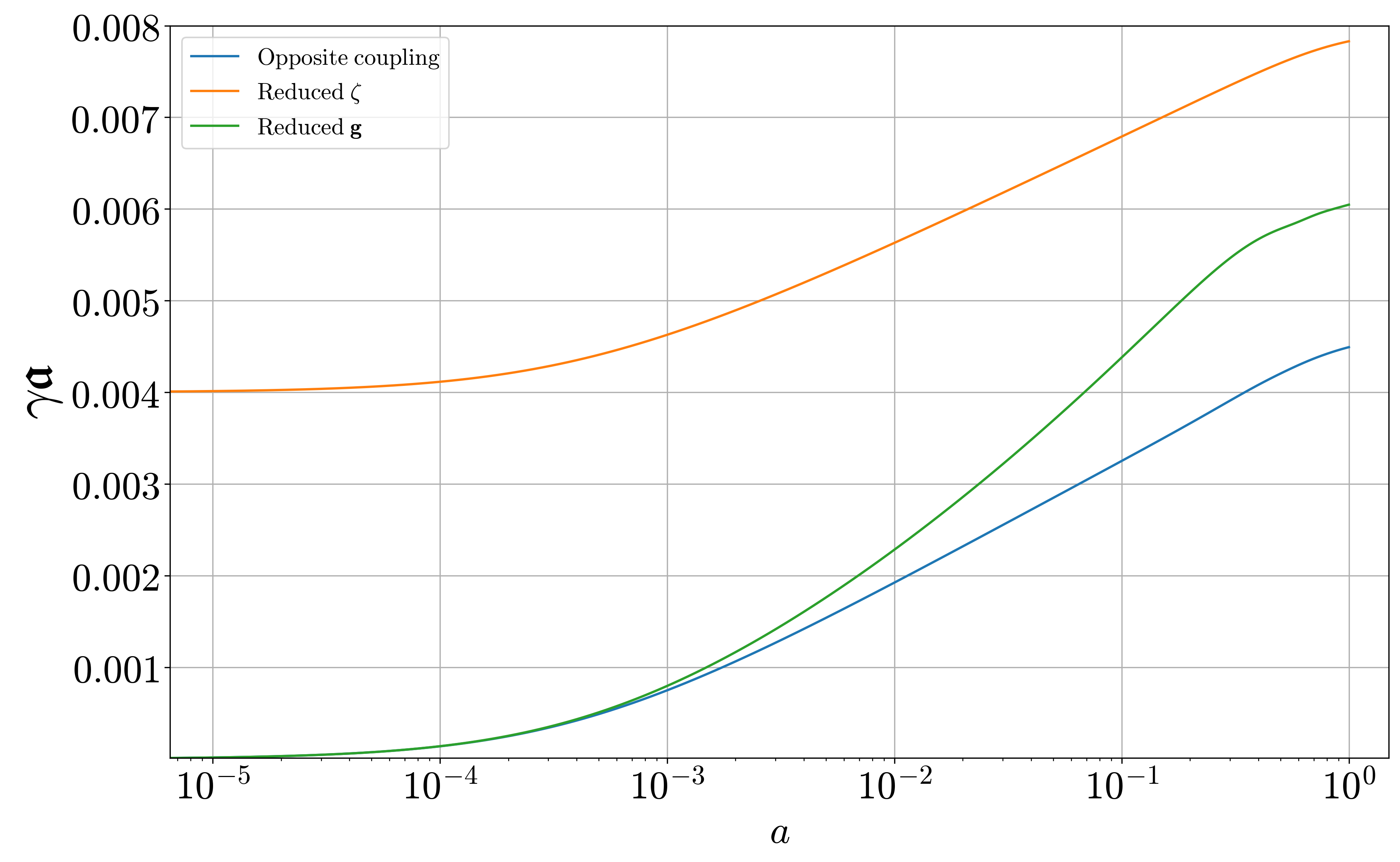}
     \caption{{\small Evolution of the product $\gamma\mfa$ for the three viable cases of table \ref{models} using the same parameters as were used in fig. \ref{fig:Backgrounds}. These confirm that $\gamma \mfa$ remains small. }}
     \label{fig:axion evol}

\end{figure}

We next turn to the remaining three cases, whose fluid energy densities are plotted in fig. \ref{fig:Backgrounds}, and suggest potentially viable cosmologies are possible. We also plot in fig. \ref{fig:axion evol} the evolution of the axion-baryon coupling strength $\gamma\mfa$ in all three of these scenarios to ensure the linear coupling approximation we use remains viable. This figure shows that the coupling term remains small in all three cases, justifying our use of the linear coupling approximation.

\subsubsection{Yoga Models with modified dilaton-DM couplings}\label{Yoga Models: dilaton differentiation of DM Section}

While the previous example shows how large matter-dilaton couplings can obstruct being able to find viable cosmological evolution, the next example shows it need not be a deal breaker. This second model differs from the Universal Yoga model only by making different assumptions about how the dilaton couples to Dark Matter (Model 2 of Table \ref{models}). Having Dark Matter couple differently than ordinary matter is quite possible within the Yoga assumptions, since these are mute about what the Dark Matter candidate is. 

We here explore the case where ordinary matter couples with Yoga-motivated strength, $\bfg_\ssB = - \frac12 \, \zeta$ and $\zeta = \sqrt{\frac23}$, but where the Dark Matter coupling is varied seeking an acceptable cosmology. We find that satisfying cosmological tests can be possible if we choose $\mathbf{g}_\CDM$ opposite in sign to $\mathbf{g} _\ssB$ and approximately an order of magnitude smaller. For example, the numerical evaluation of the cosmic fluid densities is shown in the top row of fig.~\ref{fig:Backgrounds} for the choice $\mathbf{g}_\CDM \simeq -\frac15 \, \mathbf{g} _\ssB$, and shows how the dilaton is completely stabilized and dominated by potential energy as required to describe the Dark Energy.

In this case stabilization is possible because the opposite sign dilaton coupling allows the dilaton's interaction with the Dark Matter density to drive $\chi$ towards smaller values long enough to allow it to be trapped at the minimum of $V_{\rm Yoga}$. A slightly smaller coupling suffices due to the larger density available in Dark Matter relative to baryons. 

As also noted in \cite{Burgess:2021obw}, it is noteworthy that these cosmologies robustly predict that particle masses generically differ near recombination relative to their present-day values, and moreover they do so even if they are chosen with initial conditions at nucleosynthesis that precisely agree with present-day values. This occurs because the underlying scale invariance of the field equations implies the dilaton field $\chi$ very generically falls into a scaling solution of its field equations in which its energy density likes to scale with the dominant energy density in the universe at any given time. As a result it very generically goes through an excursion around matter-radiation equality as the dominant type of tracker solution changes. 

\begin{figure}[hbtp!]
    \centering
     \begin{subfigure}[b]{0.98\textwidth}
         \centering
         \includegraphics[width=\textwidth]{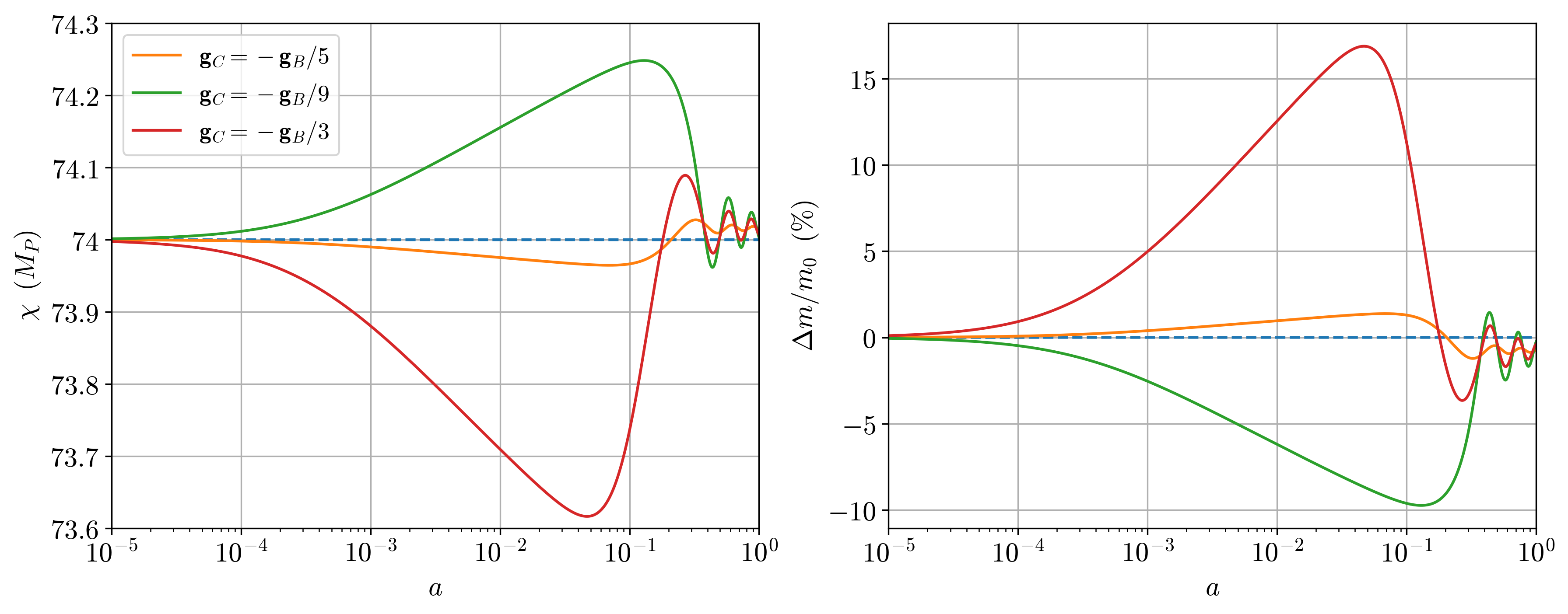}
     \end{subfigure}
     \hfill
    \begin{subfigure}[b]{0.98\textwidth}
         \centering
         \includegraphics[width=\textwidth]{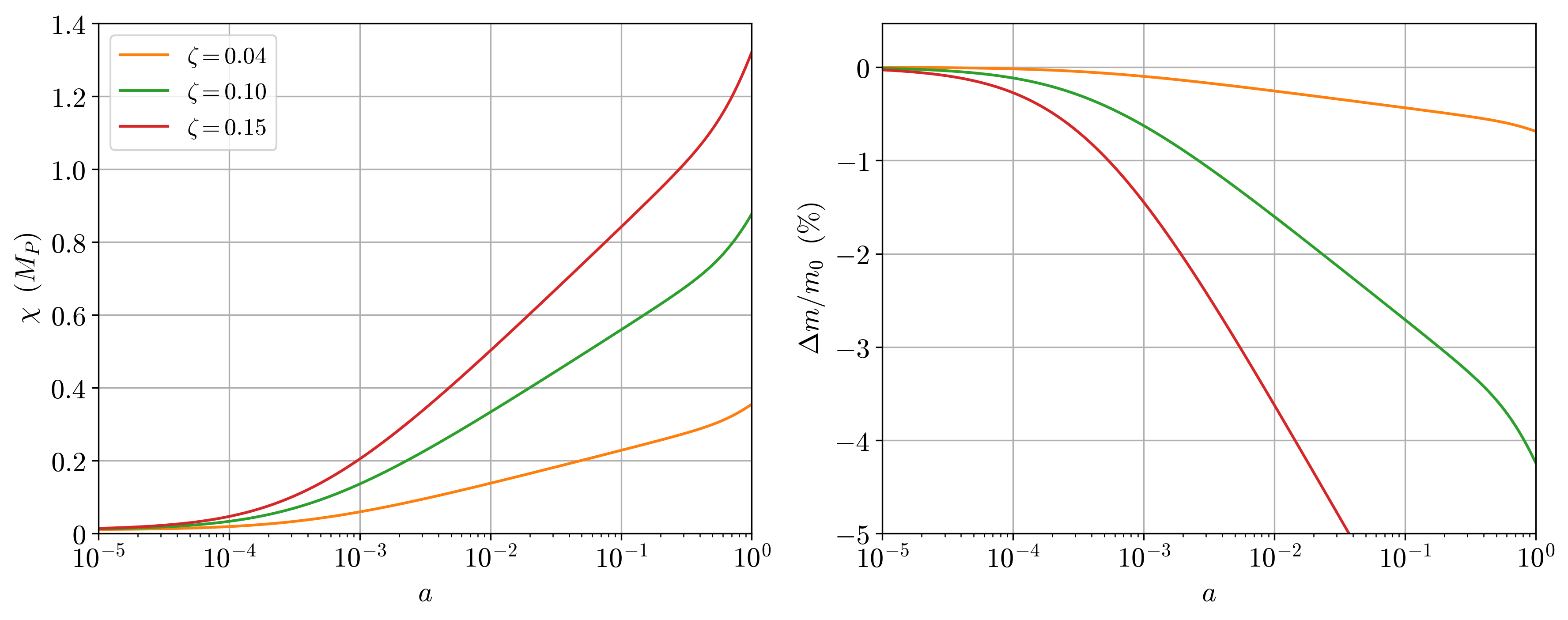}
     \end{subfigure}
    \hfill
    \begin{subfigure}[b]{0.98\textwidth}
         \centering
         \includegraphics[width=\textwidth]{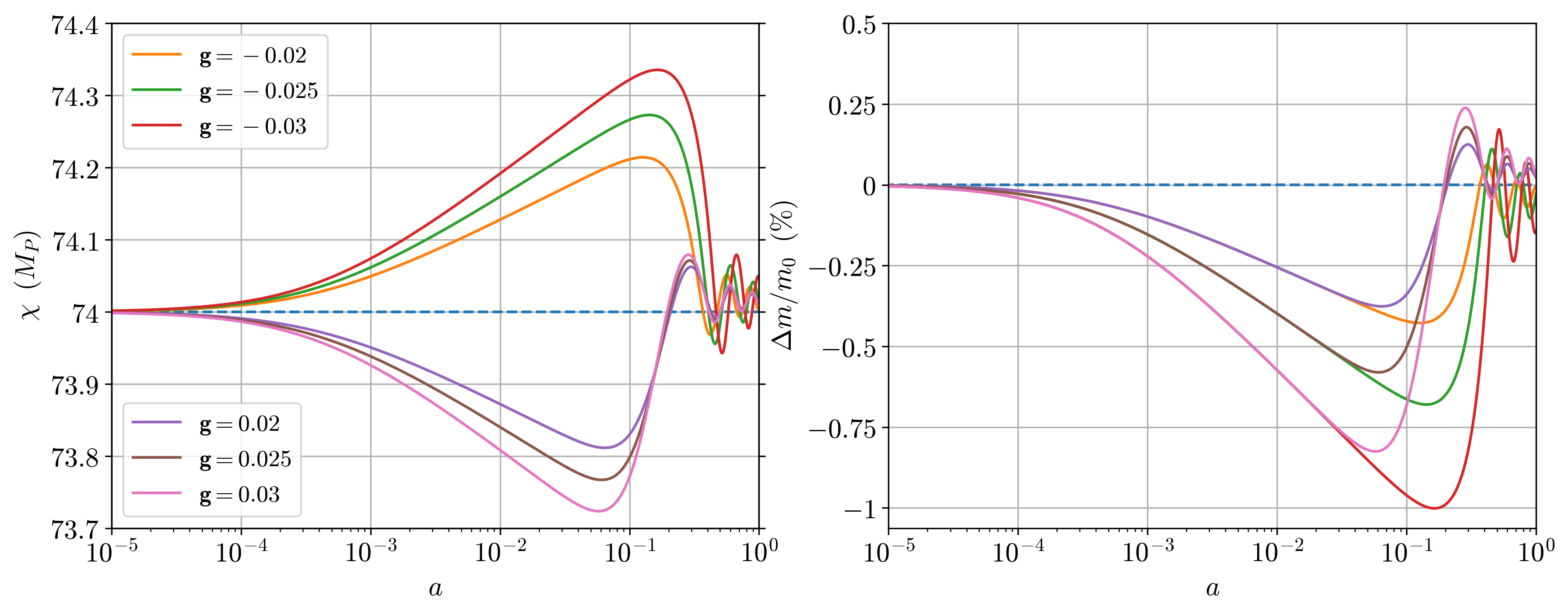}
     \end{subfigure}
     
    \caption{{\small Evolution of dilaton field and percentage change in baryon masses for the three viable cases in table \ref{models}. Top row: Model 2 with couplings $\mathbf{g} _\ssB  =-\zeta/2$ and $\mathbf{g_\CDM} = -\mathbf{g} _\ssB /5, \;-\mathbf{g} _\ssB /9,\;-\mathbf{g} _\ssB /3$ in orange, green and red respectively with $\gamma = -8\times10^{-29}$ and $\zeta = \sqrt{\frac23}$. Middle row: Model 3 with couplings $\mathbf{g} \equiv \mathbf{g} _\ssB  = \mathbf{g}_\CDM = -\zeta/2$ and $\zeta = 0.04, \;0.1, \;0.15$ in orange, green and red respectively with $\gamma = -4\times10^{-2}$. Bottom row: Model 4 with couplings $\mathbf{g}\equiv\mathbf{g} _\ssB  = \mathbf{g}_\CDM = -0.02, -0.025, -0.03$ in orange, green and red respectively and for $\mathbf{g} = 0.02, 0.025, 0.03$ in purple, brown and pink respectively with $\gamma = -8\times10^{-29}$ and $\zeta = \sqrt{\frac23}$.}}
        \label{fig:dilaton evol and ISW effect}
\end{figure}

The resulting evolution is shown in the first row of panels in Fig.~\ref{fig:dilaton evol and ISW effect}, where it can be seen that the sign of the particle mass shift depends on the magnitude of the dilaton coupling to Dark Matter. This is of interest once we explore how fluctuations evolve in the next section because it provides a potential dynamical realization of mechanisms like those proposed in \cite{Sekiguchi:2020teg} for which the Hubble tension is reduced by raising the electron mass by several percent (which makes the precise epoch of recombination occur a little earlier).

\subsubsection{Weakly coupled $SL(2,\mathbb{R})$ axio-dilatons}\label{weakly coupled axiodilatons section}

The remaining two benchmark models involve the Angle-Saxion $SL(2,\mathbb{R})$-invariant axio-dilaton framework but take a more straightforward approach to suppressing the dilaton runaway: reducing the size of the dilaton-matter couplings. We consider here two possible scenarios. In one (Model 3) we keep $\mathbf{g} = \frac12 \, \zeta$ for both Dark Matter and baryons but treat the size of $\zeta$ as a free parameter. The other scenario (Model 4) involves breaking the relationship between $\bfg$ and $\zeta$ and instead regarding $\mathbf{g}$ as a free parameter while holding $\zeta = \sqrt{\frac23}$ -- and so the strength of the axio-dilaton coupling -- fixed. In both cases the weaker dilaton-matter couplings allow the dilaton to remain at smaller values of $\chi$ long enough to become trapped at the potential's minimum and stabilize the dilaton similar to what happens in axion-free models.

\medskip\noindent
\emph{Model 3: Preserving $\bfg := \bfg_\ssB = \bfg_\ssC = \frac12 \, \zeta$ and varying $\zeta$} 

\medskip\noindent
Re-scaling $\zeta$ modifies all things dilaton related. For smaller $\lambda = 4\zeta$ the dilaton's potential becomes less steep as do the dilaton-matter couplings and the axio-dilaton sigma-model interactions. The dynamics of the small-$\zeta$ limit resembles the behaviour of the single-field quintessence models studied in \cite{Albrecht_2002}. 

Small $\zeta$ also allows another simplification. In this regime the prefactor $U$ of the dilaton potential can be taken to be constant because  an adequate Dark Energy equation of state becomes possible as the dilaton rolls down the exponential hill. So, in this case we simplify the dilaton's potential to be 
\begin{equation}
    V(\chi) = V_0 \, e^{-4\zeta\chi} \, ,
\end{equation}
where $V_0$ is a constant. For constant $U$ it is always possible to shift $\chi$ by a constant and compensate for this by scaling $V_0$ and all masses appropriately.  

The second row of fig.~\ref{fig:Backgrounds} shows the background cosmology in such a scenario. The shown solutions use $\zeta = 0.04$ with $\mathbf{g}=\mathbf{g}_\ssB = \mathbf{g}_\CDM =\zeta/2$. It turns out that reducing $\zeta$ by a factor of $\sim 20$ relative to $\cO(1)$ suffices to produce a cosmology extremely similar to $\Lambda$CDM, including for fluctuations. At the background level dilaton runaway can be avoided for $\zeta \lsim 0.1$ before the dilaton's potential becomes steep enough and the matter coupling becomes strong enough to ruin agreement with the Dark Energy equation of state $\omega_\chi \simeq -1$. For $\zeta\gtrsim 0.5$ the dilaton is drawn into the  scaling solutions that stop it dominating at late times and being a good model of Dark Energy. Notice that all of these values are larger than the value $\bfg \lsim 10^{-3}$ required in the absence of screening by solar system tests of gravity.

The effects of $\chi$ not having a minimum for $V$ can be seen in the second row of fig.~\ref{fig:dilaton evol and ISW effect}, which shows how $\chi$ (and particle masses) vary over cosmological times. The absence of a minimum removes the oscillatory behaviour at late times that arises as $\chi$ gets trapped by the local minimum. It also shows that mass evolution falls below the percent level for $\zeta \lsim 0.04$. It would be worthwhile to explore more systematically how strongly such evolution can be bounded by late-time observations (like quasar spectral lines).  

\medskip\noindent
\emph{Model 4: Holding $\zeta$ fixed and reducing $\bfg$}

\medskip\noindent
Our final Yoga variant fixes $\zeta = \sqrt{\frac23}$ and achieves realistic cosmological background evolution promoting $\mathbf{g} := \mathbf{g} _\ssB =\mathbf{g}_\CDM$ to a free parameter. The third row of fig.~\ref{fig:Backgrounds} shows the background evolution found when choosing $\mathbf{g} = -0.03$ and reducing $\mathbf{g_\mfa}$ accordingly. In this case, the modifications to the Universal Yoga Model are minimal and the dilaton's potential well acts as a stabilization mechanism allowing us to achieve slightly larger values of $\mathbf{g}$ than in the previous case without spoiling the background evolution. 

In the end, for both scenarios (small $\bfg = \frac12 \,\zeta$ and small $\bfg$ but fixed $\zeta = \sqrt{\frac23}$) we find cosmology leads to similar constraints on the upper size allowed for $\bfg$ (both of which are less restrictive than would be the solar-system constraint $\bfg \lsim 10^{-3}$ in the absence of screening). 

Both of these scenarios also predict the evolution of particle masses near recombination relative to their present-day values, as can be seen for Model 4 from the second column of panels in fig.~\ref{fig:dilaton evol and ISW effect}. In this case particle masses turn out to be reduced for all of our parameter choices, because although the sign of $\chi$ depends on the sign of $\bfg$ the mass shift only sees the product $\bfg \chi$.

\subsection{CMB anisotropies and matter power spectra}

\begin{figure}[hbtp]
    \centering
     \begin{subfigure}[b]{\textwidth}
         \centering
         \includegraphics[width=\textwidth]{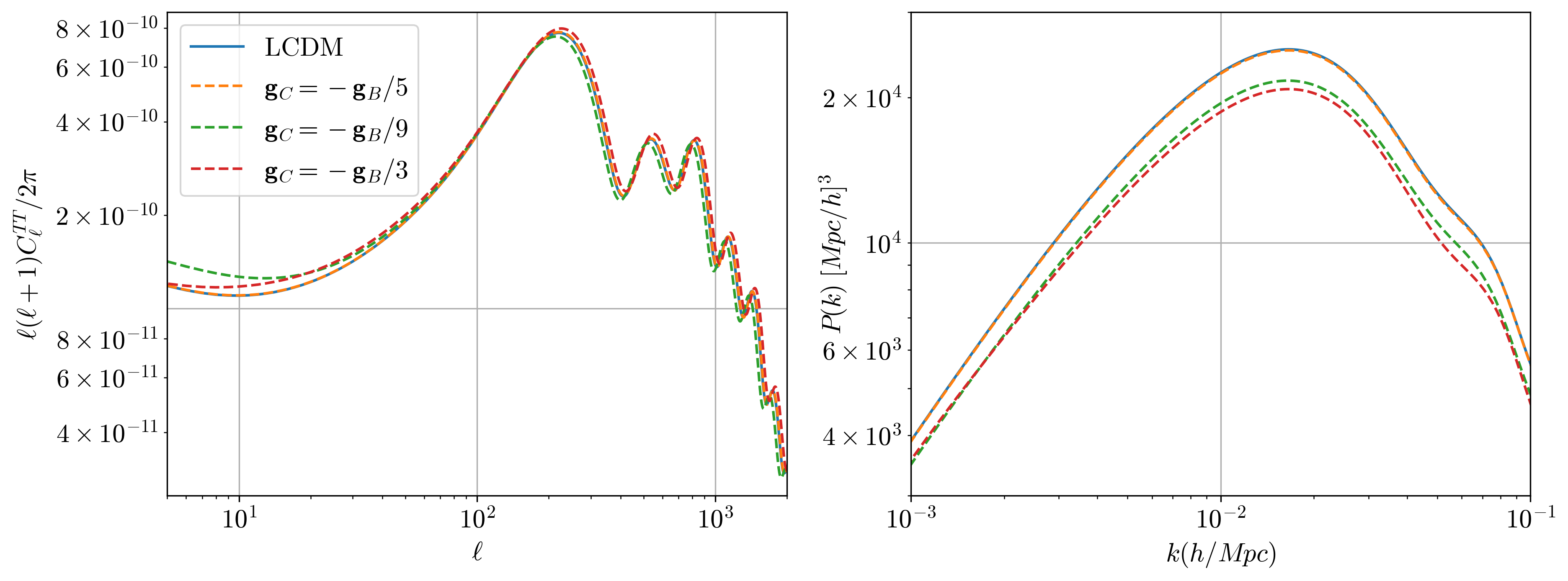}
     \end{subfigure}
     
     \hfill
     \begin{subfigure}[b]{\textwidth}
         \centering
         \includegraphics[width=\textwidth]{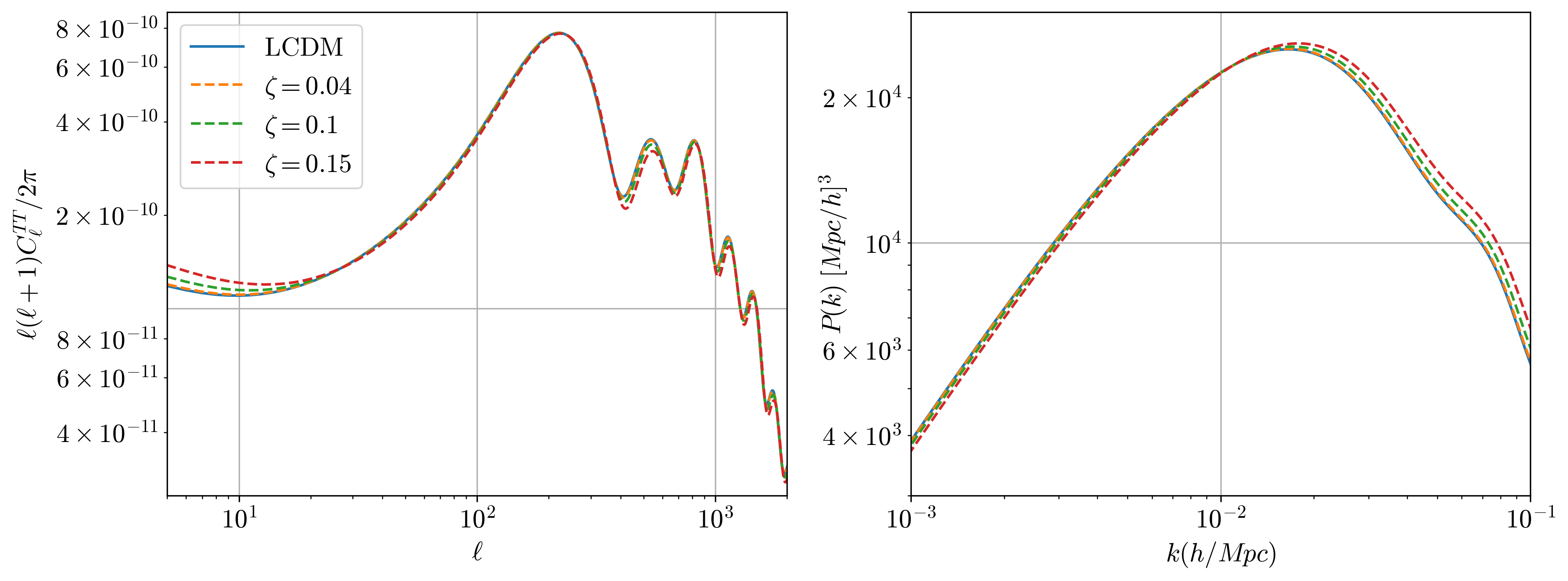}
         
         \label{fig:five over x2}
     \end{subfigure}
     
     \hfill
    
     \begin{subfigure}[b]{\textwidth}
         \centering
         \includegraphics[width=\textwidth]{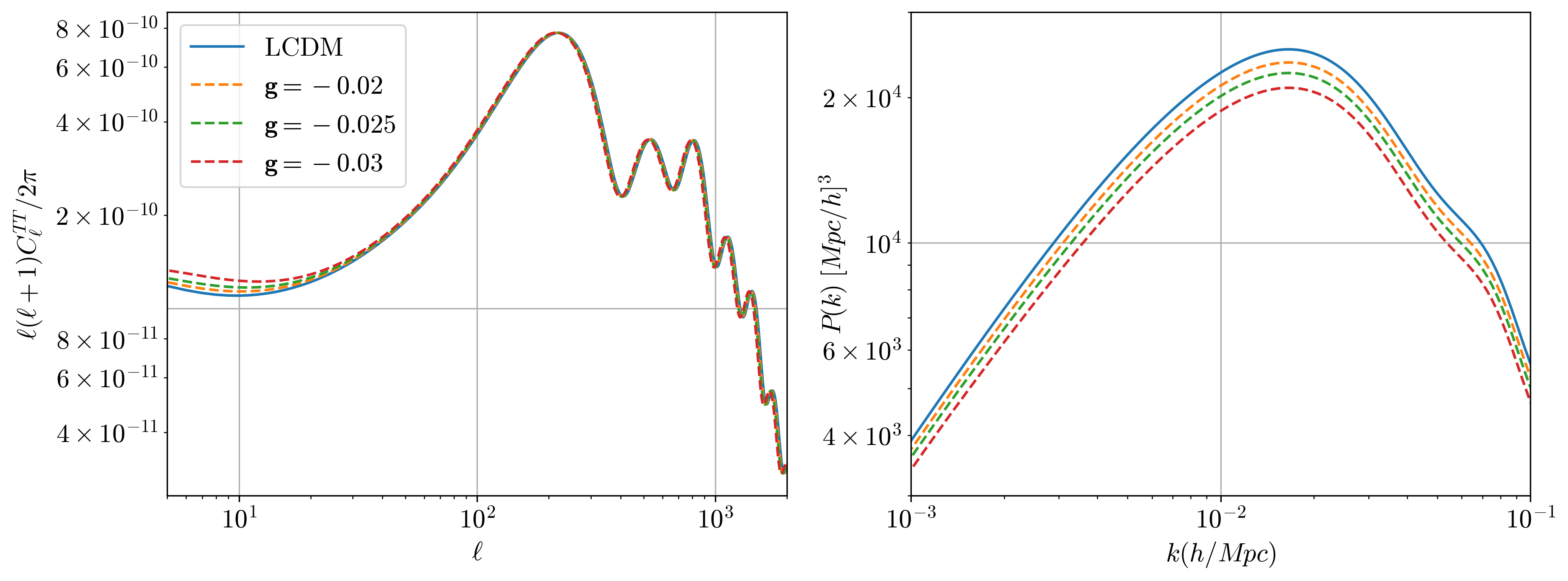}
     \end{subfigure}
    \caption{{\small Angular and matter power spectra for the three different viable cases in table \ref{models}. Top row: Model 2 with couplings $\mathbf{g} _\ssB  =-\zeta/2$ and $\mathbf{g_\CDM} = -\mathbf{g} _\ssB /5, \;-\mathbf{g} _\ssB /9,\;-\mathbf{g} _\ssB /3$ in orange, green and red respectively with $\gamma = -8\times10^{-29}$ and $\zeta = \sqrt{\frac23}$. Middle row: Model 3 with couplings $\mathbf{g} \equiv \mathbf{g} _\ssB  = \mathbf{g}_\CDM = -\zeta/2$ and $\zeta = 0.04$, $\zeta = 0.1$, $\zeta = 0.15$ in orange, green and red respectively with $\gamma = -4\times10^{-2}$. Bottom row: Model 4 with $\mathbf{g} = -0.02, -0.025, -0.03$ in dashed orange, green, red respectively with $\gamma = -8\times10^{-29}$ and $\zeta = \sqrt{\frac23}$.}}
        \label{fig:Perturbations}
\end{figure}  

We next explore the evolution of cosmological fluctuations for the three benchmark scenarios that allow viable background cosmologies. The relatively large dilaton-matter coupling strengths used in our benchmark scenarios lead to a number of important effects with implications for cosmological perturbations:

\begin{itemize}
    \item Particle masses are generically field-dependent (in Planck units). Although the Dark Matter mass depends on the dilaton field only, Standard Model particle masses (like the electron) depend in principle on both the dilaton and the axion. Even if the dilaton is ultimately trapped at the minimum of its potential, its generic excursion around recombination means that its oscillatory approach to this minimum continues to relatively late times in potentially observable ways (such as for quasar spectra). 
    
    \item Mass evolution implies the CDM energy density no longer falls inversely with universal volume, instead $\rho_m \propto e^{\mathbf{g}\chi}/a^3$. This has implications for structure formation because gravitational potentials are no longer constant during matter domination. Their evolution can affect the Integrated Sachs--Wolfe (ISW) contribution to the CMB anisotropy power spectrum. The degree to which the potentials change and hence the severity of this departure from $\Lambda$CDM is proportional to the dilaton-matter coupling strength.
    
    \item The electron mass depends dominantly on the dilaton field and the resulting time evolution changes the Thomson cross section and modifies when recombination happens. Depending on the sign of the effect this can either delay recombination or cause it to happen earlier, shifting CMB peaks relative to the best fit $\Lambda$CDM model.

    \item Although the ratio of Standard Model masses is field independent the ratio of Dark Matter to ordinary masses can vary if $\bfg_\ssC \neq \bfg_\ssB$ (or to the extent that axion evolution is important). This can allow the ratio $\rho_{\rm baryons}/\rho_{\rm CDM}$ to differ at recombination relative to today by an amount dependent on the difference $\bfg_\ssC - \bfg_\ssB$ or on the axion coupling strength. 
    
\end{itemize}

All of these effects can cause the CMB to deviate visibly relative to $\Lambda$CDM, even when the coupling $\textbf{g}$ is an order of magnitude below the value required for successful background cosmology. Fig.~\ref{fig:Perturbations} shows the results for the CMB and the matter power spectrum obtained by numerically evolving fluctuations within the three types of viable benchmark axio-dilaton cosmologies (Models 2 through 4) of Table \ref{models}. These plots confirm that a viable picture of fluctuations requires smaller couplings than are required simply for a consistent background cosmology (though larger than the $10^{-3}$ required by evasion of solar-system tests of gravity). We now highlight the specific features of each case and discuss how they differ.

\subsubsection*{Model 2: Yoga Models with modified dilaton-DM couplings}

This scenario allows the dilaton to couple more strongly to baryons than do the other two scenarios, meaning the deviations from $\Lambda$CDM associated with the changing of baryon masses are more important. This can be seen from the large deviations in baryon masses in the first row of fig.~\ref{fig:dilaton evol and ISW effect}. Varying $\mathbf{g} _\ssB $ too far on either side of $\mathbf{g}_\CDM \approx -\mathbf{g} _\ssB /5$ allows for excursions of the dilaton to a new effective minimum during matter domination, with significant implications for the evolution of cosmological perturbations. The top row of fig.~\ref{fig:Perturbations} shows the imprint on both the CMB anisotropy power spectrum and the matter power spectrum in this scenario.
 
This figure shows in particular how different CDM coupling strengths can have large effects on the second and third peaks of the angular power spectrum. This happens because the CDM coupling acts as a dilaton stabilization mechanism within its local well, with increasingly eccentric oscillations arising as the CDM coupling moves away from $\mathbf{g}_\CDM \approx -\mathbf{g} _\ssB /5$. These oscillations then lead to oscillations in the electron mass around recombination, and so locally change the time of recombination. Similar effects are observed in \cite{refId0} and can be used to place tight constraints on deviations away from $\mathbf{g}_\CDM \approx -\mathbf{g} _\ssB /5$.

\subsubsection*{Model 3: Preserving $\bfg := \bfg_\ssB = \bfg_\ssC = \frac12 \, \zeta$ and varying $\zeta$}\label{reduced zeta section}

As previously mentioned, in this case it is most economical to take $U$ to be constant and simply allow the dilaton to evolve down its exponential potential. In this scenario $\zeta$ controls both the slope of the potential and the strength of the matter coupling, and so increasing $\zeta$ causes the dilaton to roll faster, eventually producing an unacceptable dark energy equation of state with $\omega_\chi$ too far from $-1$. For example, when $\zeta \approx 0.1$ one finds $\omega_{(\chi 0 )}\approx 0.96$, and this produces significant deviations in the angular and matter power spectrum from $\Lambda$CDM, as shown in the middle row of fig.~\ref{fig:Perturbations}. Removing the quadratic minimum also removes the mechanism forcing particle masses back to their pre-recombination values at late times. As the dilaton rolls down it's exponential between matter domination and today particle masses will evolve with it, as shown in the middle row of fig. \ref{fig:dilaton evol and ISW effect}. We can see that taking $\zeta = 0.1$ results in baryon masses being reduced by $\sim4\%$. Although this places additional constraints on the size of $\zeta$ to stop baryon masses varying too much between recombination and today for a pure exponential dilaton potential, small enough $\zeta$ can produce acceptably small mass variations which could be used as a prediction in the search for late time stringy physics. 

\subsubsection*{Model 4: Holding $\zeta$ fixed and reducing $\bfg$}

Although $\mathbf{g} = -0.05$ is enough to achieve an acceptable background cosmology, slightly smaller values for $\mathbf{g}$ are required to obtain acceptable CMB ansiotropy and matter power spectrum. This can be seen in the lower row of fig.~\ref{fig:Perturbations} with the model being almost indistinguishable from $\Lambda$CDM when $|\mathbf{g}|<0.02$.
 
Fig.~\ref{fig:dilaton evol and ISW effect} shows the evolution of the dilaton field and the associated deviations in the masses of standard model particles for this scenario. These deviations are largest just after matter-radiation equality, when the matter coupling term in \ref{friedmann dilaton eval} first dominates over the Hubble friction term. In the reduced $\mathbf{g}$ case, this has a smaller effect on the angular power spectrum than the models with Yoga strength baryon couplings as the dilaton's oscillations within its potential well are much lower in amplitude. Although such reduced oscillations might help evade late-time constraints ({\it e.g.}~from quasar spectra) they also suppress the effects that could have interesting implications for the resolution of the Hubble tension, such as discussed in \cite{Sekiguchi:2020teg, Schoneberg:2021qvd}.

\subsection{A closer look at structure growth}

We close this section with a discussion of the linearised equations governing the density contrast of baryons and cold dark matter at subhorizon scales within the axio-dilaton framework. 

In the quasistatic limit the equations in (\ref{Delta B general}) and (\ref{delta C general}) simplify to
\begin{eqnarray} \label{delta B yoga}
&& \delta_\ssB''+\delta_B'\left[\mathcal{H}+\mathbf{g} _\ssB \bar{\chi}'+\mathbf{g}_\mfa\bar{\mfa}'\right] \\
&&\qquad\qquad\qquad = 4\pi a^2G_{\rm N}\bigg[\left(1+2 \, \mathbf{g} _\ssB ^2\frac{k_{ph}^2}{k_\chi^2}+2\frac{\mathbf{g}^2_\mfa}{W^2}\right)\bar{\rho}_B\delta_\ssB  +\left(1+2 \, \mathbf{g} _\ssB \mathbf{g}_\CDM\frac{k_{ph}^2}{k_\chi^2}\right)\bar{\rho}_{C}\delta_{\ssC}\bigg] \, ,\nn
\end{eqnarray}
and
\begin{equation} \label{delta C yoga}
\delta_{\CDM}''+\delta_{\CDM}'\left[\mathcal{H}+\mathbf{g}_\CDM\bar{\chi}'\right] =  4\pi a^2G_{\rm N}\left[\left(1+2 \, \mathbf{g} _\ssB \mathbf{g}_\CDM\frac{k_{ph}^2}{k_\chi^2}\right)\bar{\rho}_B\delta_\ssB + \left(1+2 \, \mathbf{g}_\CDM^2\frac{k_{ph}^2}{k_\chi^2}\right)\bar{\rho}_{\ssC}\delta_{\ssC}\right] \, ,
\end{equation}
where $k_{ph} = k/a$, $k_\chi^2 = \frac{k^2}{a^2}+m_\chi^2$ and $G_{\rm N}$ is Newton's gravitational constant. These equations capture the linearized physics of structure formation in the sub-horizon regime. 

In particular, these equations highlight the impact of coupling the axion to baryons but not to CDM. For baryons both the dilaton and the axion mediate a fifth force interaction that affects growth while only the dilaton does so for Dark Matter, with implications for the evolution of the baryon density contrast relative to CDM.  As a result we expect different growth rates for density fluctuations of baryons and cold dark matter. Furthermore, (\ref{delta B yoga}) and (\ref{delta C yoga}) show that the masslessness of the axion field implies the axion-baryon modification is present across all length scales. The dilaton-matter modification, by contrast, is suppressed on length scales larger than its Compton wavelength: $m_\chi^2\gg \frac{k^2}{a^2}$. The axion contribution is important because its occurs with Planck strength ({\it i.e.}~$F \simeq M_p$ for the parameters used in the plots) while the dilaton-matter coupling is suppressed by $\mathbf{g}^2$ relative to gravitational strength.

\begin{figure}[hbt!]
    \begin{subfigure}[b]{\textwidth}
        \includegraphics[width=\textwidth]{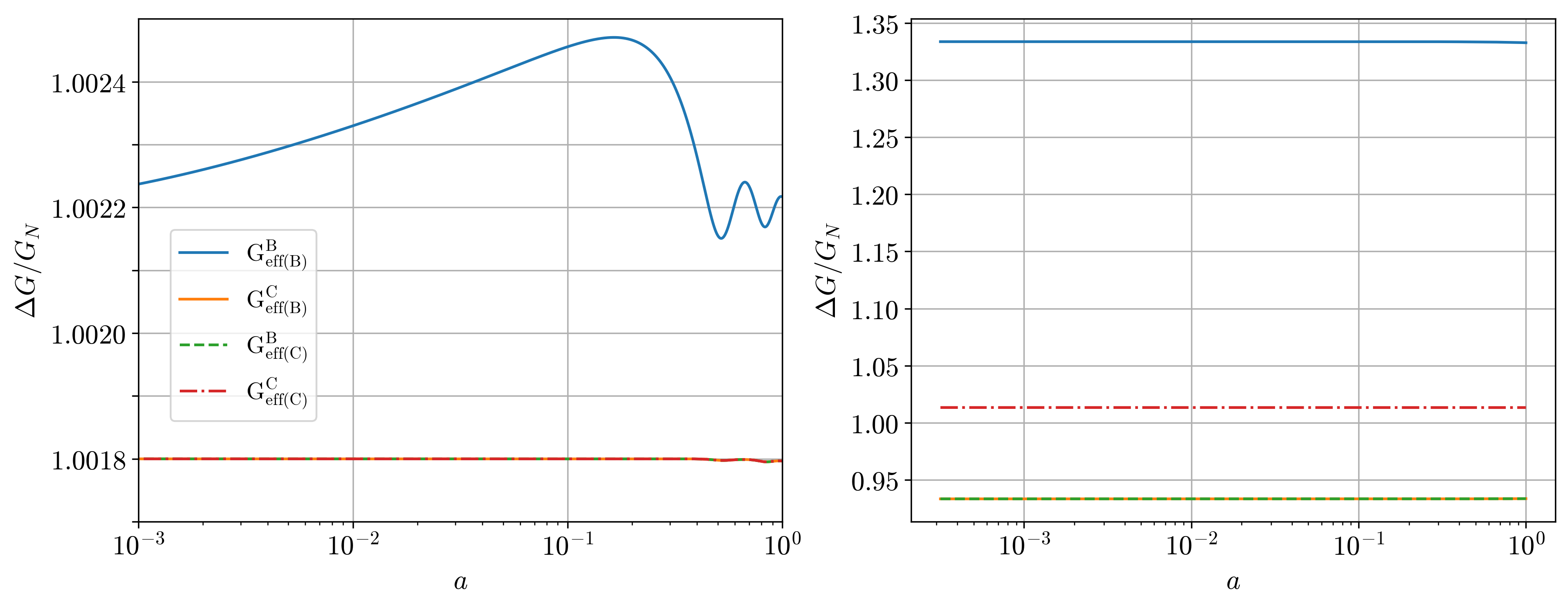}
    \end{subfigure}

    \caption{{\small Evolution of the deviations of the effective gravitational couplings from Newton's constant for baryon-baryon (blue), CDM-CDM (red), baryon-CDM (orange), CDM-baryon (green dashed) interactions for $k = 0.125h\rm\; M_{pc}^{-1}$. On the left we have chosen $\mathbf{g}_{\ssC} = \mathbf{g} _\ssB  = -0.03$, whereas on the right we set $\mathbf{g}_{\ssC} = -\mathbf{g} _\ssB /5$ and $\mathbf{g} _\ssB  = -\zeta/2$. For both cases we set $\gamma = -8\times 10^{-29}$.}}
    \label{fig:newtons consts}
\end{figure}

The evolution of the effective Newton's constants in (\ref{delta B yoga}) and (\ref{delta C yoga}) is shown in fig.~\ref{fig:newtons consts} for two of the models considered above. This shows that even in the reduced coupling case -- {\it i.e.}~when $\mathbf{g} _\ssB  = \mathbf{g}_\CDM$ -- the additional axion clumping term gives rise to an extra contribution to baryon-baryon clustering, provided the axion coupling strength is large enough to compete with the dilaton's. For the Yoga model with opposite sign dilaton-CDM and dilaton-baryon couplings, this effect is overshadowed by the extremely strong baryon coupling required by the Yoga model, which increases the strength of baryon-baryon and CDM-CDM clustering while diminishing baryon-CDM interactions. The example shown in fig.~\ref{fig:newtons consts} uses $\mathbf{g}_{\CDM} = -\mathbf{g} _\ssB /5$, and in this case it is interesting to note that increasing the baryon-baryon clustering strength by $\sim 30\%$ results in negligible deviations from $\Lambda$CDM in both the angular and matter power spectra, as can be seen in fig.~\ref{fig:Perturbations}.

The second terms on the left-hand side of (\ref{delta B yoga} and \ref{delta C yoga})  depict the modified Hubble friction, associated with the slowing of the growth of structure formation due to the expansion of the universe. When the net velocity of the fields is positive, their coupling to the matter species reduces these friction effects, as seen with the axion, thereby boosting the growth of baryonic structures even more. In a typical single-field dark energy model with an exponential potential one would expect that these adjustments (the increased effective Newton's constant and the decreased Hubble friction terms) would accelerate the early stages of structure formation, as observed in \cite{Brax:2023tls}. The introduction of a potential well for the dilaton, however, turns out to drastically change the dynamics of structure growth. In such a scenario, the dilaton's restricted evolution  within the well stops any net contributions from reducing the Hubble friction. Additionally, its displacement from the minimum of its well during matter domination, shown in fig.~\ref{fig:dilaton evol and ISW effect}, acts to increase the Hubble rate. Both of these factors can actually result in decreased structure growth.

\begin{figure}[hbt!]
    \begin{subfigure}[b]{\textwidth}
        \centering
        \includegraphics[width=\textwidth]{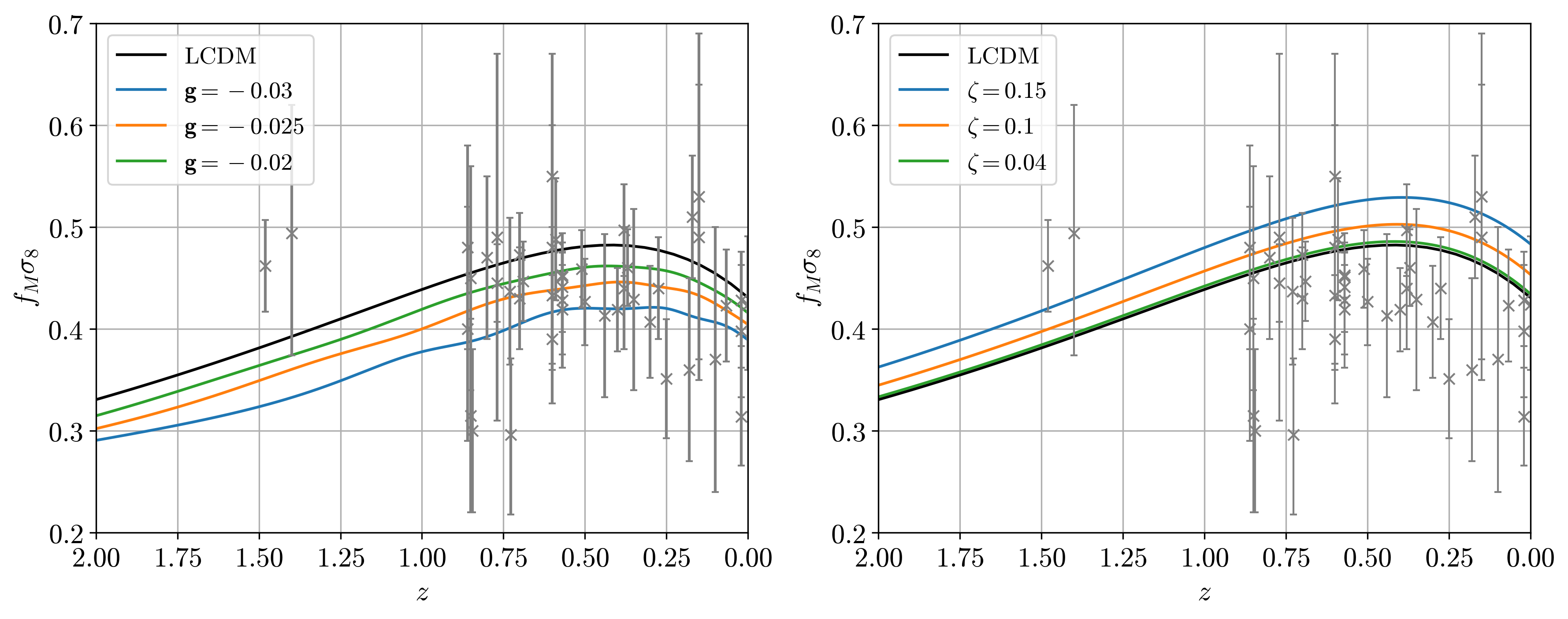}
        
    \end{subfigure}
    
    \caption{{\small Here we show $f{\sigma_8}$ for the General dilaton-matter coupling (Model 3, on the left) and the Reduced $\zeta$ (Model 4, on the right) cases.
    On the left $\zeta = \sqrt{\frac23}$ and the couplings are chosen to be $\mathbf{g}\equiv \mathbf{g} _\ssB =\mathbf{g}_\CDM = -0.03,\;-0.025,\;-0.02$ in blue, green and orange respectively. On the right the couplings are $\mathbf{g}\equiv \mathbf{g} _\ssB =\mathbf{g}_\CDM = -\zeta/2$ and $\zeta = 0.15,\;0.1,\;0.04$ blue, green and orange respectively. The data points used are taken from \cite{Marulli_2021} and the black lines give the $\Lambda$CDM prediction.}}
    \label{fig:fsigma8}
\end{figure}

We can parameterise these effects using the linear growth rate of each species, defined as 
\begin{equation}
    f_i(z, k) := \frac{1}{\mathcal{H}}\frac{\delta_i'(z,k)}{\delta_i(z,k)} \, .
\end{equation}
A quantity sensitive to this parameter that can be directly probed by observations of redshift space distortions \cite{Song:2008qt} is given by
\begin{equation}
    f\sigma_8 = \frac{\sigma_8(z,k_{\sigma8})}{\mathcal{H}}\frac{\delta_m'(z,k_{\sigma8})}{\delta_m(z,k_{\sigma8})} \, ,
\end{equation}
where $k_{\sigma8} = 0.125h~$Mpc$^{-1}$, $\delta_m = (\delta \rho_\ssB +\delta \rho_\CDM)/(\rho_B+\rho_C)$ and $\sigma_8$ is the variance of the mass fluctuations within a sphere of radius $R=8 h^{-1}$Mpc, defined by
\begin{equation}
   \sigma_8^2 = \int \frac{dk}{k} |{\cal W}(kR)|^2 \Delta^2(k) \,,
\end{equation}
where ${\cal W}(kR)$ is the Fourier transform of the real--space top-hat window function and $\Delta^2(k)$ is the dimensionless power spectrum defined by $\Delta^2(k) = k^3P(k)/2\pi^2$. 

Fig.~\ref{fig:fsigma8} shows the evolution of this quantity between matter domination and today for the reduced $\mathbf{g}$ and reduced $\zeta$ benchmark scenarios (Models 3 and 4). In contrast to the reduced-$\zeta$ case (Model 3) -- which takes the dilaton's potential to be a pure exponential -- the existence of the potential well in the reduced $\mathbf{g}$ case (Model 4) ensures that larger coupling strengths correspond to smaller $f\sigma_8$ at late times. This is because larger couplings result in larger excursions of the dilaton from the minimum of its potential, as shown in fig.~\ref{fig:dilaton evol and ISW effect}, providing an extra contribution to the Hubble rate and hence decreasing structure growth even more. 

This effect also leads to a reduction in the parameter $\sigma_8$ at the present epoch, independently of the linear growth rates, warranting a more thorough data analysis of dark energy models with such a potential well.

\section{Quadratic axion-matter couplings}
\label{QuadraticAxionCouplingSection}

In this section we finish our analysis with a preliminary discussion of how the above cosmologies change in the presence of an axion scalar potential and if the function $\cU(\mfa)$ that controls the axion-dependence of particle masses has a local minimum. We show here that these additions seem to be consistent with successful cosmology.

\subsection{Motivations and benchmark model}

These are interesting modifications to the axio-dilaton models for many reasons, not least because within matter such a choice can cause $\mfa$ to be dynamically driven to a point where $\cU'$ (and so also $\ga$) vanishes. This kind of dynamics can dramatically change the strength of axion constraints (such as by reducing the axion emission rate within hot stars \cite{Beadle:2023flm}).   

Such modifications are also motivated by the solar-system sized elephant in the room: axio-dilaton models with Brans-Dicke couplings $\bfg_\ssB \gsim 10^{-3}$ are at face value ruled out by solar system tests of gravity. This need not be an issue for either of the low-coupling Angle-Saxion scenarios described above, for which $\bfg_\ssB \lsim 10^{-3}$ can be chosen consistent with successful cosmology and the axion does not play an important role in stabilizing the dilaton. But it is potentially a problem for the Yoga-style models with larger dilaton-matter couplings. Such models only make sense for cosmology within the context of some sort of screening mechanism that allows them to evade non-cosmological constraints.  

Screening mechanisms are relatively poorly explored for multiple-scalar models like the axio-dilatons of interest here, but one was recently proposed in ref.~\cite{Brax:2023qyp}. This mechanism relies on three ingredients: ($i$) there is an axion contribution $V(\mfa)$ to the scalar potential; ($ii$) the function $\cU(\mfa)$ has a local minimum; and ($iii$) the functions $V(\mfa)$ and $\cU(\mfa)$ are minimized for different values $\mfa_\pm$ of the axion field. With these ingredients parameters can be chosen so that the axion energy is minimized for different field values inside and outside of matter, which sets up gradients in the axion near a macroscopic object's surface. These gradients then work through the derivative axion-dilaton interactions to suppress the object's coupling to the dilaton. 

This section presents a preliminary attempt to see whether these ingredients can be consistent with sensible cosmological evolution. We explore how the previously described cosmology might change in the presence of quadratic axion-matter couplings and a vacuum axion potential.

We take the same basic setup as for the Yoga model and for concreteness do so with the best-case opposite-coupling scenario (Model 2) described in \ref{Yoga Models: dilaton differentiation of DM Section} with $\mathbf{g}_\CDM = -\mathbf{g} _\ssB /5$. The scalar potential is the sum of the dilaton potential $V_{\rm Yoga}$ given in \pref{YogaPotential} and a vacuum axion potential\footnote{Although easier to calculate with, strictly speaking this choice for scalar potential strays from the Yoga program for which axion dependence would actually enter as an $\mfa$-dependence of $U$ appearing in $V_{\rm Yoga}$ rather than as a separate additive term (whose zero value at the minimum is not explained).} $V(\mfa)$. The axion-matter coupling $\ga$ is then modified just enough to allow the function $\cU$ to have a nontrivial minimum. 

Although we know that both $V(\mfa)$ and $\cU(\mfa)$ are likely to be periodic functions within any real microscopic formulation, we expect the axion dynamics to be largely dominated by evolution near their minima and so for simplicity assume the easy-to-compute-with form
\begin{equation} \label{galin}
    \ga = \hat \gamma^2 (\mfa- \mfa_-)   \qq{and} V(\mfa) = \frac{1}{2} \mu_\mfa^2 \MPL^2 (\mfa-\mfa_+)^2 \, ,
\end{equation}
and we choose $\mfa_- <  \mfa_+$. Although this expression for $\ga$ is convenient for numerical calculations it is also deceptively simple. Recalling $\ga = \cU'/\cU$ it corresponds to choosing the function $\cU(\mfa)$ to be
\begin{equation}
    \cU(\mfa) = \exp\left[ \frac12 \,\hat \gamma^2 (\mfa- \mfa_-)^2 \right]  \simeq 1 + \frac12 \, \hat \gamma^2 (\mfa - \mfa_-)^2 + \cdots \, ,
\end{equation}
where the approximate equality holds only if $\hat \gamma |\mfa - \mfa_-| \ll 1$. In order for \pref{galin} to mimic an approximately quadratic function for $\cU(\mfa)$ we must check that $\hat \gamma(\mfa - \mfa_-)$ never gets too large.

If the axion is canonically normalized for a specific choice of background value $\bar\chi$ the parameter $\mu_a$ corresponds to a vacuum axion mass 
\begin{equation}
    m_\mfa = \frac{\mu_\mfa \MPL}{f} = \frac{\mu_\mfa}{W(\bar\chi)} \,.
\end{equation}
The coupling $\hat\gamma$ similarly defines a new scale $\Lambda_\mfa = \MPL/\hat\gamma$ in terms of which the axion-matter coupling is characterized by a `decay constant' 
\begin{equation}
    \hat F := \frac{f}{\hat \gamma} = \frac{W(\bar\chi) \MPL}{\hat \gamma}= W(\bar\chi) \Lambda_\mfa \,.
\end{equation}
In the presence of matter the axion field moves as if in a matter-dependent effective potential, $V_{\rm eff}(\mfa,\rho_\ssB)$, given explicitly by \pref{VeffMat}. This implies the axion also has a matter-dependent mass (compare to \pref{meffdef})
\begin{equation} \label{meffdef2}
    W^2  m_{\rm eff}^2 (\mfa, \rho_\ssB) = \frac{1}{\MPL^2} \left[  V_{\mfa\mfa}(\mfa) + \left( \frac{\cU_{\mfa\mfa}}{\cU} \right)\, \rho_\ssB  \right] \nn\\
    =  \mu_\mfa^2 + \Bigl[ 1 + \hat \gamma^2 (\mfa - \mfa_-)^2 \Bigr] \frac{\rho_\ssB}{\Lambda_\mfa^2} \,,
\end{equation} 
and so $m_{\rm eff}^2 \simeq m_\mfa^2 + (\rho_\ssB/\hat F^2)$ in the regime $\hat \gamma (\mfa - \mfa_-) \ll 1$.

Motivated by the discussion in \cite{Brax:2023qyp} we choose benchmark values $\mu_\mfa \Lambda_a = m_\mfa \hat F \sim 1 \; \rm eV^2$, chosen so that the density where the two terms in \pref{meffdef2} compete occurs not too far above cosmologically interesting densities and not too from the surface of terrestrial and solar-system objects. (For the purposes of comparison notice that in these units terrestrial objects have an average density of $\rho_{\rm ter}\sim 1 \; \hbox{g/cm}^2 \sim 4 \times 10^{18} \; \rm eV^4$ while the current cosmic energy is of order $\rho_{\rm vac} 
\sim  10^{-3}\rm eV^4$.)
When needed we also choose $m_\mfa \sim 10^{-15}$ eV and $\hat F \sim 10^6$ GeV in order to arrange present-day screening depths to be conveniently smaller than solar-system objects. See \cite{Brax:2023qyp} for a discussion of particle-physics constraints on $\hat F$.

\subsection{Cosmological evolution}

From the cosmological point of view these choices imply $m_\mfa \gg \mathcal{H}$ and so the axion field oscillates around the minimum of its potential with a frequency much faster than the Hubble rate. Such quick oscillations are difficult to simulate numerically over cosmological timescales and so we sidestep this by using the Madelung formalism for a scalar field fluid (\cite{Madelung:1927ksh}; see e.g. \cite{Ferreira:2020fam} for a summary and review in the context of scalar field DM and \cite{Brax_2019} for its use in the case of self-interacting DM). 

This approach assumes the axion field begins its evolution sufficiently close to the minimum of $V_{\rm eff}(\mfa, \rho_\ssB)$ and splits the axion field into an adiabatic part, $\bar \mfa(\rho_m)$, that tracks the minimum of the effective potential, plus a rapidly oscillating part describing the fast oscillations around this value:
\begin{equation}\label{axion split}
    \mfa = \bar\mfa(\rho_\ssB) + \frac{1}{\sqrt 2}\left[ e^{-i\int_0^t dt\, m(t)}\psi + e^{i\int_0^t dt\, m(t)}\psi^\star\right] \, ,
\end{equation}
where factoring out the oscillatory factor means $\psi$ evolves much more slowly. $m(t)$ here denotes $m_{\rm eff}$ as given in \pref{meffdef2}.
 
The dynamics of the background axion field is given by solving $V_{\rm eff}'(\mfa,\rho_\ssB) = 0$ and so
\begin{equation}\label{bg ax fluid evol}
    \bar\mfa = \frac{\mu^2_\mfa \MPL^2 \mfa_+ + \hat \gamma^2 \rho_\ssB \, \mfa_-}{\mu^2_\mfa \MPL^2 + \hat \gamma^2 \rho_\ssB} \, .
\end{equation}
%
%
%
The evolution of the slowly evolving field $\psi(\bfx,t)$ defines an effective axion fluid density, as is most easily seen by re-expressing it in terms of its modulus and phase:
\begin{equation}\label{madelung density}
    \psi = \frac{\sqrt{\rho_\mfa}}{W(\Bar{\chi})m_{\rm eff}} \; e^{iS} \, ,
\end{equation}
where $S(\bfx,t)$ denotes the phase and the modulus is written so that $\rho_{\mfa}(\bfx,t)$ is related to the physical axion energy density -- {\it i.e.}~the $00$ component of the axion energy--momentum tensor -- by $T^{00} = \frac{1}{2} \rho_{\mfa}\left[1 + \frac{m_{\mfa}^2}{m_a^2(\rho_m)} \right]$. 

Defining the axion fluid velocity by $\vec v_a \equiv { \nabla S}/[{m(t) a}]$, energy conservation for $\psi$ can be obtained by substituting \pref{axion split} and \pref{madelung density} into the axion Lagrangian, leading to the fluid-like equation
\begin{equation}
    \rho_\mfa'+3{\cal H}\rho_\mfa - \frac{m'}{m}\rho_\mfa + (\nabla \cdot \vec{v}_\mfa) \rho_a = 0 \, ,
\end{equation}
where again $m(t) = m_{\rm eff}(\rho_\ssB)$. At the background level this has the solution
\begin{equation}
    \Bar{\rho}_\mfa = \frac{C m_{\rm eff}(\rho_\ssB)}{a^3} \, ,
\end{equation}
where $C$ is an integration constant determined by the initial axion fluid energy density.  

In cosmological evolution our numerical parameter choices ensure the mass $m_{\rm eff}(\rho_\ssB)$ approaches the vacuum value $m_a$ at sufficiently late times and so the axion fluid evolves as $\rho_\mfa \propto 1/a^3$, behaving like dust in the usual way. At earlier times the higher baryon density eventually implies $m^2_{\rm eff} \propto \rho_\ssB$ and so $\rho_\mfa \propto \sqrt{\rho_\ssB}/a^3 \propto1/a^{9/2}$. We conclude from this that because the axion fluid falls more quickly than radiation in the early universe that its share of the universal energy density is diluted, ensuring it is subdominant for late time cosmology. For the present analysis we therefore set $C = 0$ and disregard the axion fluid altogether. (A more complete treatment of this fluid is performed in \cite{screenedcosmo}.) A similar story applies to $\psi$ fluctuations: the super-Hubble value for the local mass at the minimum of the effective potential ensures that these perturbations quickly decay away, justifying their neglect.

\begin{figure}[hbt!]
    \centering
     \begin{subfigure}[b]{\textwidth}
         \centering
         \includegraphics[width=\textwidth]{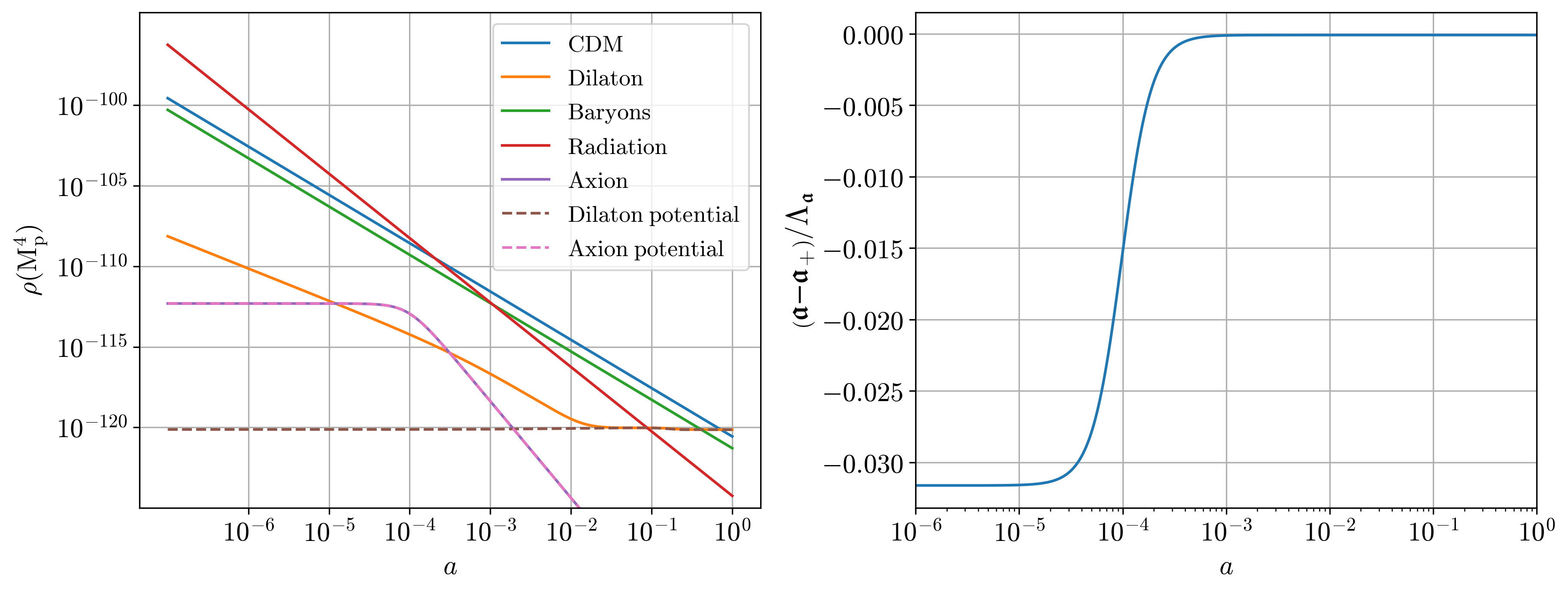}
     \end{subfigure}
     \hfill
     
     \begin{subfigure}[b]{\textwidth}
         \centering
         \includegraphics[width=\textwidth]{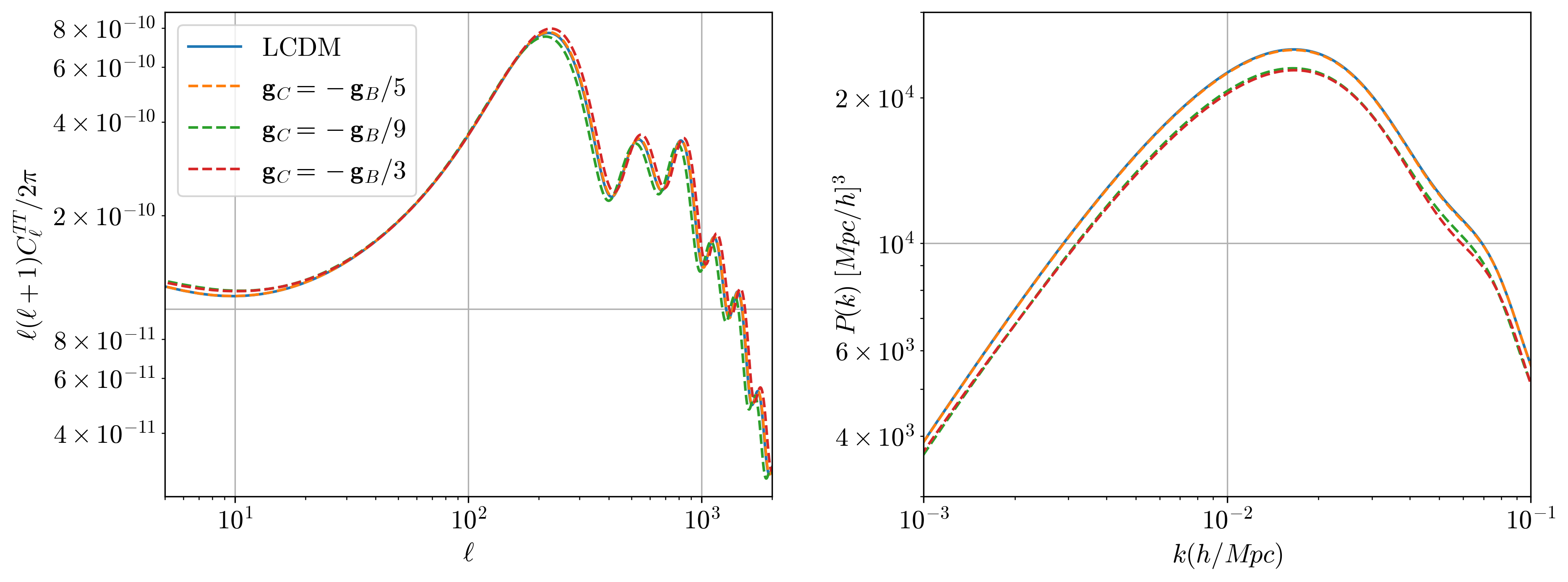}

     \end{subfigure}
    \hfill
    \caption{{\small Top left shows the background evolution of the energy density in baryons (green), CDM (blue), radiation (red) dilaton total (orange), dilaton potential (brown dashed) axion field total (purple), axion field potential (dashed pink). Top right shows the normalised axion field evolution. Bottom row depicts the dimensionless angular and matter power spectra for different couplings compared to the LCDM best fit in blue. In the background plots $\mathbf{g}_{C} = -\mathbf{g}_B/5 = \frac{1}{5\sqrt{6}}$, and for all plots $m_{\mfa} = 4\times10^{-15} \rm eV$, $\Lambda_a = 5\times10^{5} \rm GeV$, and $\mfa_+ - \mfa_- = 2\times10^4 \rm GeV$.}}
        \label{fig:quadratic background}
\end{figure}

Under these circumstances the axion field can simply be replaced by $\bar \mfa$. The remaining dilaton and Friedmann equations then become
\begin{align}\label{friedmann dilaton ax fluid}
    \Bar{\chi}''+2\mathcal{H}\Bar{\chi}'+\zeta e^{-2\zeta\chi}\Bar{\mathbf{a}}'^2+\frac{a^2}{\MPL ^2}\left[V,_\mathbf{\chi}(\Bar{\chi}, \Bar{\mathbf{a}}\right)+\mathbf{g} _\ssB \Bar{\rho} _\ssB +\mathbf{g}_\CDM\Bar{\rho}_\CDM]=0 \, ,
    \\
    \mathcal{H}^2 = \frac{1}{3\MPL ^2}\left[\left(\frac{\Bar{\chi}'^2}{2}+W^2\frac{\Bar{\mathbf{a}}'^2}{2}\right)\MPL ^2+a^2V(\Bar{\chi},\Bar{\mathbf{a}})+a^2\Bar{\rho}\right] \,,
\end{align}
and it is these equations that we integrate numerically to evolve the background and cosmological fluctuations, for the opposite coupling Yoga model with $\mathbf{g}_\CDM = -\mathbf{g} _\ssB /5 = \frac{1}{5\sqrt{6}}$. Fig.~\ref{fig:quadratic background} shows the result of such a calculation, showing the evolution of the energy density of each fluid as well as the computed form for the CMB angular distributions and the power spectrum. We see from the figure that sub-dominant nature of the axion ensures the deviations from the best fit $\Lambda$CDM are minimal within this scenario, for both background evolution and for the angular and matter power spectrum. Note that the axion evolution (shown in the top-right panel) is small, justifying the use of a quadratic expansion around the minima of the potential and coupling. 

Although these results are only preliminary they are encouraging inasmuch as they suggest that axio-dilaton screening mechanisms like that discussed in \cite{Brax:2023qyp} can be embedded into the large-coupling Yoga model in a way that is consistent with both solar system and cosmological observations.

\section{Conclusions and outlook}
\label{Conclusions}

In this paper we provided a detailed study of the cosmological implications of axio-dilaton scalar-tensor theories. We analysed the cosmological background evolution from deep inside the radiation dominated epoch to the present day and presented, for the first time for such models, the results for the CMB anisotropy power spectrum and the matter power spectrum. In doing so, we focused on four cases, summarized in Table \ref{models}. The models differ by the size of the couplings of the dilaton and axion to matter (collectively denoted by {\bf g}, ${\bf g}_B$ and ${\bf g}_{\it C}$) and the choice of the kinetic coupling parameter $\zeta$. We find that the universal Yoga model, formulated with equal DM and baryon couplings, is most likely not viable because of two conflicting constraints. Either the axion-matter coupling is weak and the large dilaton coupling to matter prevents the dilaton from finding the minimum of its potential. But if the axion coupling is made larger then the large axion-matter coupling causes the baryon mass to evolve too much. 

We find two categories of theories that lead to a more realistic background evolution. In one the dilaton couples to matter with large Yoga-sized strength, but the couplings to baryons and CDM (that is ${\bf g}_B$ and ${\bf g}_{\it C}$) are not equal. The second category of viable model allows the couplings to be universal, but smaller than in the universal Yoga model. Such models can also lead to realistic predictions for the CMB anisotropy and matter power spectra. 

Going beyond these minimal four benchmark models, we also present a preliminary discussion of a model in which the axion potential is not vanishing and the axion coupling function has a local minimum. We find that such a model also can lead to a viable cosmology and our results provides a new avenue for model building. They also suggest exploring more systematically how strongly mass evolution can be bounded by late-time observations.  

Although our calculations find viable cosmologies, viability depends on using a particular type of initial condition (at least for those models where the dilaton potential has a minimum). We assume that the initial value of the dilaton deep in the radiation epoch is not far from its present-day value, since having particle masses too different in the past could ruin the successes of Big Bang nucleosynthesis. A more complete model might hope to explain why the dilaton should start off this way. (Interestingly there is not also a strong restriction on the dilaton's initial velocity at these early times because there can be ample time for Hubble friction to drain this away.) Because the present-day field lies close to the potential minimum, and the initial dilaton field cannot be too far away from there, that overshoot of the minimum can become a problem when dilaton-matter couplings are too strong. A proper theory of initial conditions might require information about the UV completion and/or any earlier inflationary history (about which we remain agnostic in this paper). 

These models suggest a number of interesting directions that remain to be explored. One such asks whether the axion alone can be the DM if it is given an appropriate potential. Such a framework would seek to build an axionic DM model out of the axiodilaton's axion, explaining DM as arising from the oscillations of the axion field around the minimum of its potential. The question is whether having the axion couple to baryons and the dilaton as envisaged here ruins the success of such models. Can standard production mechanisms for axionic DM be incorporated? How would the predictions for CMB anisotropies and the matter power spectrum be different from the cases we have explored here? We intend to address some of these issues in a future publication \cite{axionCDM}. 

Another interesting extension to the work presented here would conduct a full MCMC data analysis of the viable scenarios presented here. Although this goes beyond the scope of the present work, the dynamical ability to alter particle masses described in these models could be relevant to mechanisms aimed at various cosmological tensions. These include the role of the electron mass variations proposed to alleviate the Hubble tension \cite{Schoneberg:2021qvd}, and the role of the dilaton oscillating in its dark energy potential well in reducing structure growth rates at late times. Because axio-dilaton theories are simple and string-friendly they provide a framework for introducing both late time and early time new physics through more complicated choices in the axio-dilaton potential and couplings than those studied here. 

Axio-dilaton scalar-tensor theories also provide a rich phenomenology for the very early universe, where they have long been studied in the context of embedding inflationary models into string theory (for a review see \cite{Cicoli:2023opf}). Having multiple scalar fields play important cosmological roles also suggests more novel cosmological epochs, some of which are explored in \cite{Cicoli:2023opf,Apers:2024ffe}, such as by allowing post-inflationary epochs of kination, tracker and moduli-domination that can considerably delay the onset of radiation domination even just before big bang nucleosynthesis. 

In real compactifications (including modulus-stabilization mechanisms - for a review see \cite{McAllister:2023vgy}) the underlying approximate scaling symmetries often ensure the existence of tracker solutions that eventually settle into a local minimum \cite{Burgess:2016ygs, Apers:2024ffe}. Similar things occur in the specific models explored here; the relaxation field present in Yoga models is an attractive inflaton candidate \cite{Burgess:2022nbx}, which both alleviates some of the $\eta$-problems encountered by other approaches and suggests an interesting post-inflationary history.

Having common ingredients at both early and late times might suggest new mechanisms for solving old problems and new observational windows for exploring the much earlier universe. A good place to start are the minimal models, such as extending the late-universe axion-dilaton system described here into the very early universe in more detail to understand if it can give rise to interesting modifications to early-universe dynamics.

\section*{Acknowledgements}

We thank Adam Solomon for helpful conversations and Elsa Teixeira for the useful numerical resources. This work evolved out of discussions at the Astroparticle Symposium at the Institut Pascal. CB, CvdB and ACD thank the Institut Pascal for their hospitality during the programme. MM is also grateful for the hospitality of Perimeter Institute where part of this work was carried out. AS is supported by the W.D. Collins Scholarship. CvdB is supported by the Lancaster–Sheffield Consortium for Fundamental Physics under STFC grant: ST/X000621/1. ACD is partially supported by the Science and Technology Facilities Council (STFC) through the STFC consolidated grant ST/T000694/1. CB's research was partially supported by funds from the Natural Sciences and Engineering Research Council (NSERC) of Canada. MM is supported in parts by the Mid-Career Research Program (2019R1A2C2085023) through the National Research Foundation of Korea Research Grants. MM is also supported Kavli IPMU which was established by the World Premier International Research Center Initiative (WPI), MEXT, Japan. This work was also supported by a grant from the Simons Foundation (1034867, Dittrich). Research at the Perimeter Institute is supported in part by the Government of Canada through NSERC and by the Province of Ontario through MRI. 

\appendix

\bibliographystyle{JHEP}
\bibliography{bibliography}

\end{document}